\documentclass[journal]{IEEEtran}
\PassOptionsToPackage{hyphens}{url}
\usepackage{hyperref}
\usepackage{graphicx}
\usepackage{footnote}
\usepackage{url}
\usepackage[table,xcdraw]{xcolor}
\usepackage[caption=false,font=footnotesize]{subfig}
\usepackage{multirow}
\usepackage{adjustbox}
\usepackage{pifont}
\usepackage{amsmath}
\usepackage{marvosym}
\usepackage{booktabs}
\usepackage{threeparttable}
\usepackage{xfrac}
\usepackage{multirow}
\usepackage{pifont}
\usepackage{cite}
\usepackage{resizegather}
\usepackage{tabu}
\usepackage{algorithm}
\usepackage{algpseudocode}
\usepackage{optidef}
\usepackage{makecell}

\usepackage{acronym}
\acrodef{CMOS}[CMOS]{Complementary Metal–Oxide–Semiconductor}
\acrodef{CNN}[CNN]{Convolutional Neural Network}
\acrodef{NN}[NN]{Neural Network}
\acrodef{RNN}[RNN]{Recurrent Neural Network}
\acrodef{EEG}[EEG]{Electroencephalogram}
\acrodef{iEEG}[iEEG]{Intracranial Electroencephalography}
\acrodef{DL}[DL]{Deep Learning}
\acrodef{GPU}[GPU]{Graphic Processing Unit}
\acrodef{MAC}[MAC]{Multiply-Accumulate}
\acrodef{VMM}[VMM]{Vector-Matrix Multiplication}
\acrodef{FIR}[FIR]{Finite Impulse Response}
\acrodef{ML}[ML]{Machine Learning}
\acrodef{ADC}[ADC]{Analog-to-Digital Converter}
\acrodef{DAC}[DAC]{Digital-to-Analog Converter}
\acrodef{IoMT}[IoMT]{Internet of Medical Things}
\acrodef{MDLS}[MDLS]{Memristive Deep Learning Systems}
\acrodef{RRAM}[RRAM]{Resistive Random-Access Memory}
\acrodef{IMC}[IMC]{In-Memory Computing}
\acrodef{SVM}[SVM]{Support Vector Machine}
\acrodef{kNN}[kNN]{k-Nearest Neighbor}
\acrodef{RF}[RF]{Random Forest}
\acrodef{1T1R}[1T1R]{1-Transistor-1-Resistor}
\acrodef{IoT}[IoT]{Internet of Things}
\acrodef{AFE}[AFE]{Analog Front End}
\acrodef{FIR}[FIR]{Finite Impulse Response}
\acrodef{CA}[CA]{Cellular Automata}
\acrodef{ESN}[ESN]{Echo State Network}
\acrodef{ANN}[ANN]{Artifical Neural Network}
\acrodef{AUROC}[AUROC]{Area Under the Receiver Operating Characteristic Curve}
\acrodef{MLP}[MLP]{Multi-Layer Perceptron}
\acrodef{SNN}[SNN]{Spiking Neural Network}
\acrodef{SOP}[SOP]{Seizure Occurance Period}
\acrodef{SPH}[SPH]{Seizure Prediction Horizon}
\acrodef{RMSA}[RMSA]{Root Mean Square Amplitude}
\acrodef{PCA}[PCA]{Principal Component Analysis}
\acrodef{SVD}[SVD]{Singular Value Decomposition}
\acrodef{QAT}[QAT]{Quantization Aware Training}
\acrodef{DBS}[DBS]{Deep Brain Stimulation}
\acrodef{FFT}[FFT]{Fast Fourier Transform}
\acrodef{FPR}[FPR]{False Positive Rate}
\acrodef{FPGA}[FPGA]{Field Programmable Gate Array}
\acrodef{VLSI}[VLSI]{Very-large-scale Integration}
\acrodef{SOTA}[SOTA]{State-Of-The-Art}
\acrodef{SPICE}[SPICE]{Simulation Program with Integrated Circuit Emphasis}
\acrodef{BEOL}[BEOL]{Back-End-Of-The-Line}
\acrodef{TDM}[TDM]{Time-Division Multiplexing}
\acrodef{CAD}[CAD]{Computer Automated Design}
\acrodef{SRAM}[SRAM]{Static Random-Access Memory}
\acrodef{SNN}[SNN]{Spiking Neural Network}
\acrodef{SWEC}[SWEC]{Sleep-Wake-Epilepsy-Center}
\acrodef{DRAM}[DRAM]{Dynamic Random-Access Memory}
\acrodef{PCM}[PCM]{Phase-Change Memory}
\acrodef{MRAM}[MRAM]{Magnetoresistive Random-Access Memory}
\acrodef{CBRAM}[CBRAM]{Conductive Bridging Random-Access Memory}
\acrodef{WL}[WL]{Word Line}
\acrodef{BL}[BL]{Bit Line}

\usepackage{tikz}

\usepackage{multirow}
\usepackage{orcidlink}

\title{Seizure Detection and Prediction by Parallel Memristive Convolutional Neural Networks}

\begin{document}
	\author{
		Chenqi Li$^\dagger$,~\IEEEmembership{Student Member,~IEEE,}\thanks{\hspace{-1em}\rule{3cm}{0.5pt} \newline \textcopyright  \hspace{1pt} 2022 IEEE. Personal use of this material is permitted. Permission from IEEE must be obtained for all other uses, in any current or future media, including reprinting/republishing this material for advertising or promotional purposes, creating new collective works, for resale or redistribution to servers or lists, or reuse of any copyrighted component of this work in other works.}
		Corey Lammie$^\dagger$\orcidlink{0000-0001-5564-1356},~\IEEEmembership{Student Member,~IEEE,}
		Xuening Dong,~\IEEEmembership{Student Member,~IEEE,}
		Amirali Amirsoleimani\orcidlink{0000-0001-5760-6861},~\IEEEmembership{Member,~IEEE,}
		Mostafa Rahimi Azghadi\orcidlink{0000-0001-7975-3985},~\IEEEmembership{Senior Member,~IEEE,}
		and~Roman Genov\orcidlink{0000-0001-7506-1746},~\IEEEmembership{Senior Member,~IEEE}
		\thanks{$^\dagger$\textbf{These authors contributed equally.}
		
		\textit{Corresponding authors: M. Rahimi Azghadi and A. Amirsoleimani}.
		
		Chenqi Li,  Xuening Dong, and Roman Genov are with the Department of Electrical and Computer Engineering, University of Toronto, Toronto, Canada. email: chenqi.li@mail.utoronto.ca, xuening.dong@mail.utoronto.ca, roman@eecg.utoronto.ca
	
		Corey Lammie and M. Rahimi Azghadi are with the College of Science and Engineering, James Cook University, QLD 4811, Australia. e-mail: corey.lammie@jcu.edu.au, mostafa.rahimiazghadi@jcu.edu.au}
		
		\thanks{Amirali Amirsoleimani is with the Department of Electrical Engineering and Computer Science, York University, Toronto, Canada. e-mail: amirsol@yorku.ca}}

	\markboth{ACCEPTED BY IEEE TRANSACTIONS ON BIOMEDICAL CIRCUITS AND SYSTEMS, 2022}
	{Li, Lammie \MakeLowercase{\textit{et al.}}: Memristive Seizure Detection and Prediction}
	\maketitle

\begin{abstract}
During the past two decades, epileptic seizure detection and prediction algorithms have evolved rapidly. However, despite significant performance improvements, their hardware implementation using conventional technologies, such as \ac{CMOS}, in power and area-constrained settings remains a challenging task; especially when many recording channels are used. In this paper, we propose a novel low-latency parallel \ac{CNN} architecture that has between 2-2,800x fewer network parameters compared to \ac{SOTA} \ac{CNN} architectures and achieves 5-fold cross validation accuracy of 99.84\% for epileptic seizure detection, and 99.01\% and 97.54\% for epileptic seizure prediction, when evaluated using the University of Bonn \ac{EEG}, CHB-MIT and SWEC-ETHZ seizure datasets, respectively. We subsequently implement our network onto analog crossbar arrays comprising \ac{RRAM} devices, and provide a comprehensive benchmark by simulating, laying out, and determining hardware requirements of the \ac{CNN} component of our system. To the best of our knowledge, we are the first to parallelize the execution of convolution layer kernels on separate analog crossbars to enable 2 orders of magnitude reduction in latency compared to \ac{SOTA} hybrid Memristive-\ac{CMOS} \ac{DL} accelerators. Furthermore, we investigate the effects of non-idealities on our system and investigate \ac{QAT} to mitigate the performance degradation due to low \ac{ADC}/\ac{DAC} resolution. Finally, we propose a stuck weight offsetting methodology to mitigate performance degradation due to stuck $R_{\text{ON}}/R_{\text{OFF}}$ memristor weights, recovering up to 32\% accuracy, without requiring retraining. The \ac{CNN} component of our platform is estimated to consume approximately 2.791W of power while occupying an area of 31.255mm2 in a 22nm FDSOI CMOS process.
\end{abstract}

\begin{IEEEkeywords}
CNN, Seizure Detection, Seizure Prediction, EEG, RRAM, Memristive Crossbar Array
\end{IEEEkeywords}
\IEEEpeerreviewmaketitle
	
\section{Introduction}
\begin{figure}[!t]
    \centering
    \includegraphics[width=0.5\textwidth]{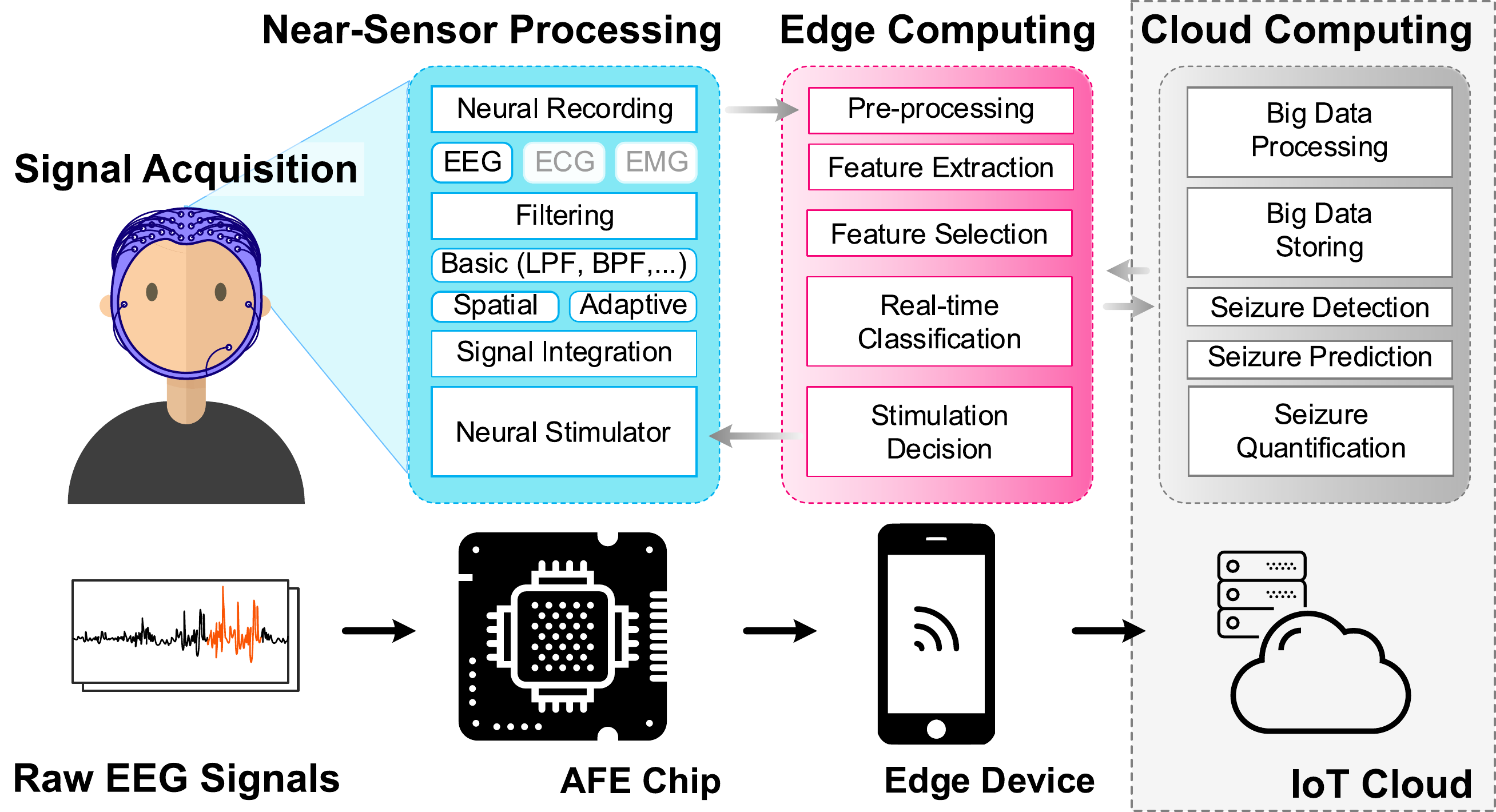}
    \caption{An overview of a typical epileptic seizure detection and prediction system. Acquired \ac{EEG} signals are sampled and processed near-sensor using an \acf{AFE}, prior to being sent wirelessly to edge device(s) for real-time pre-processing and feature extraction. Features can then be fed into \ac{ML} and/or \ac{DL} architectures, residing either on the \ac{IoT} edge or in the \ac{IoT} cloud, which perform epileptic seizure detection and prediction.
    }
    \label{fig:1}
\end{figure}

\IEEEPARstart{E}{pilepsy} is a common neurological disorder that affects approximately 1\% of the world's population~\cite{Beghi2019}. A seizure is characterized by excessive firing of neurons in the brain, while epilepsy is a medical condition that involves recurrent seizures~\cite{Stafstrom2015}. As the underlying occurrence mechanism of epilepsy is not well understood~\cite{Patel2019,Brennan2018,Gasparini2019}, it requires experimental methods of treatment that rely on accurate detection and prediction systems, as depicted in Fig.~\ref{fig:1}.

\ac{EEG} is the most common method used to monitor the electrical activities of the brain, and can be used to detect and predict seizures. There have been numerous applications of traditional \ac{ML} algorithms, such as \acp{SVM}, \ac{kNN}, and \ac{RF} classifiers to classify ictal (seizure), preictal (prior to a seizure) and non-ictal (non-seizure) signals using \ac{EEG} recordings. Despite being able to achieve high accuracies, these approaches require the manual extraction and selection of features in the time- or frequency-domain~\cite{Siddiqui2020}. The optimal choice of such feature extractions are largely unknown, experimental, and dependant on specific patient signatures, such that there is no one-fit-all solution.

Compared to traditional seizure classification algorithms, \ac{DL}-based algorithms have more advantages in complex \ac{EEG} signal feature extraction, as they do not require feature engineering, and are capable of outperforming traditional \ac{ML} algorithms for epileptic seizure detection and prediction tasks~\cite{Ibrahim2022}. However, when these \ac{DL} systems are implemented using \ac{CMOS}, there are problems such as large scale, high calculation energy consumption and high delay, which hinder their efficacy; especially in resource-constrained environments.

In order to solve this kind of problem, this paper proposes a neuromorphic calculation strategy based on a novel \ac{IMC} \ac{RRAM} architecture, which utilizes analog crossbars. Computer designers have traditionally separated the role of storage and compute units. The \ac{IMC} paradigm blurs this distinction, and imposes the dual responsibility on memory substrates: storing and computing on data for massively parallel computing~\cite{Das2022}. By exploiting the physical characteristics of emerging analog device technologies, analog crossbars can be used to perform \acp{VMM}, the most dominant operation in \acp{CNN}, in as little as $\mathcal{O}(1)$~\cite{Li2018,Lammie2019}, significantly reducing the computational complexity during inference operations. Our specific contributions are as follows:

\begin{enumerate}
    \item To the best of our knowledge, we are the first to parallelize the execution of convolution layer kernels on separate analog crossbars to address the computational bottleneck of \acp{CNN}, enabling 2 orders of magnitude reduction in latency compared to current \ac{SOTA} hybrid Memristive-\ac{CMOS} \ac{DL} accelerators;
    \item We reduce the number of required parameters by 2-1,600x and 5-2,800x for epileptic seizure detection and prediction tasks using deep learning models, while still achieving SOTA performance;
    \item We provide a comprehensive benchmark for hardware memristor-based seizure prediction/detection systems by simulating, laying out, and determining hardware requirements of the CNN component of our system;
    \item We propose a simplified stuck weight offsetting methodology for mitigating severe degradation of system performance due to stuck $R_{\text{ON}}/R_{\text{OFF}}$ memristor weights. We demonstrate that our method is capable of achieving up to 32\% performance recovery, without requiring retraining, while incurring minimal hardware and computational overhead.
\end{enumerate}
To promote reproducible research, all of our simulation codes are made publicly accessible\footnote{\url{https://github.com/coreylammie/Memristive-Seizure-Detection-and-Prediction-by-Parallel-Convolutional-Neural-Networks}}. The rest of the paper is structured as follows: In Section~\ref{sec:related_work}, we overview and discuss related work.
In Section~\ref{sec:novel_system}, we present our epileptic seizure detection and prediction system. In Section~\ref{sec:software_methodology}, we overview and discuss our software methodology. In Section~\ref{sec:hardware_simulation_methodology}, we overview and discuss our hardware simulation methodology. In Section~\ref{sec:results_discussion}, we present and discuss our results. Finally, we conclude the paper in Section~\ref{sec:conclusion}.

\begin{figure*}[!t]
     \centering
  \includegraphics[width=0.95\textwidth]{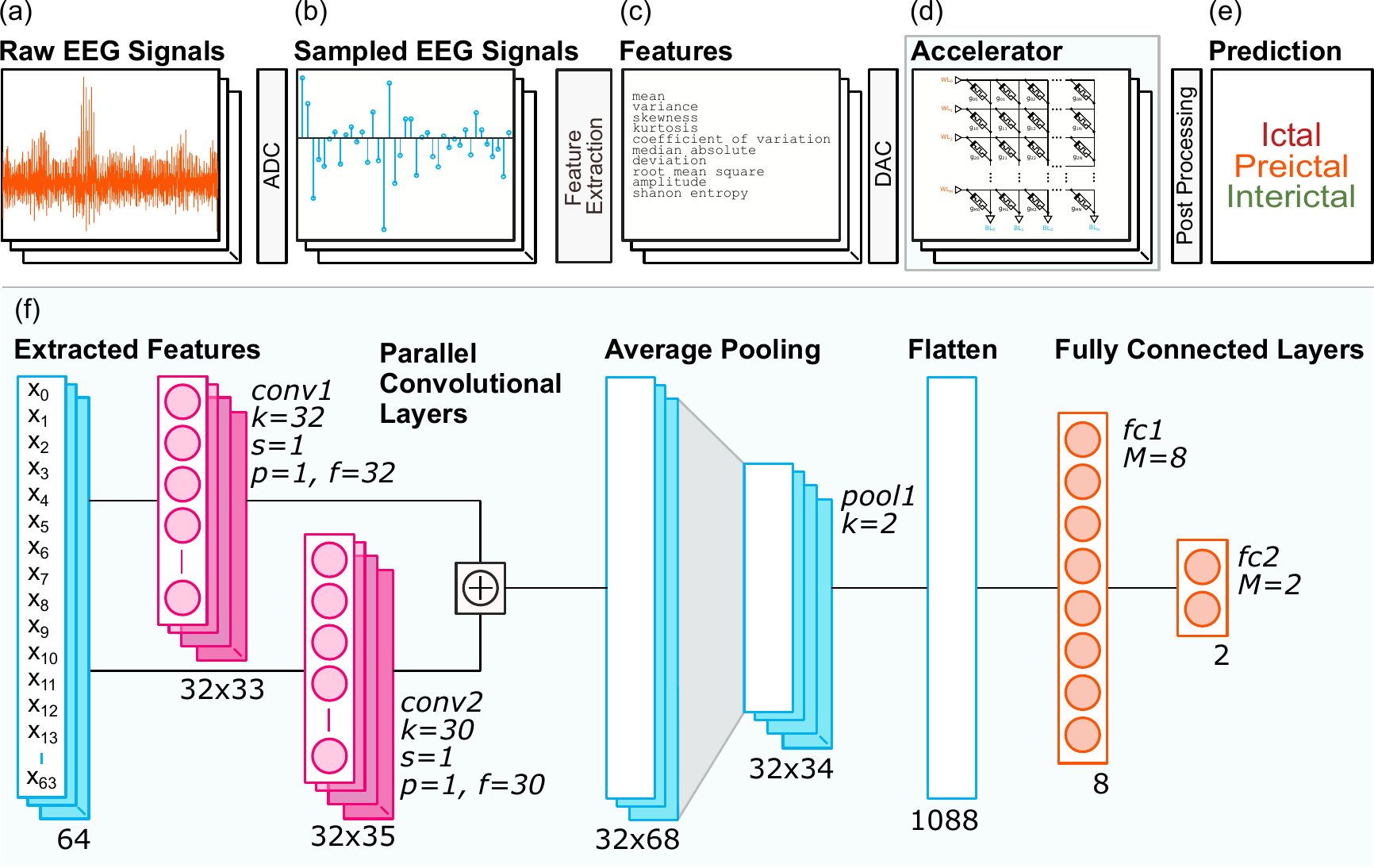}
  \caption{A high-level system architecture overview. (a) Raw \ac{EEG} signals are sampled and digitized using \acp{ADC}. (b) Features are extracted from sampled \ac{EEG} signals. (c) Extracted features are fed into a memristive \ac{DL} accelerator. (d) Accelerator outputs are processed. Fig.~\ref{fig:block_diagram} depicts the detailed hardware implementation of the accelerator. (e) Processed accelerator outputs are used to determine interictal, preictal, and ictal states. (f) The novel neural network architecture used consists of two parallel 1d-convolutional layers, one average pooling layer, and two fully connected (dense) layers. $N$ is used to denote the batch size, i.e., the number of batches presented to the network in parallel. $f$ denotes the number of filter. $k$ determines the filter size. $s$ denotes the stride length. $p$ denotes the padding. $M$ denotes the number of output neurons for each fully connected layer. Parts of this figure are derived from~\cite{Lammie2021}.}
  \label{workflowl}
\end{figure*}

\section{Related Work}\label{sec:related_work}
In this section, we present an overview of related work using parallel \acp{CNN} and related work using traditional and neuromorphic \ac{ML} architectures for the detection and prediction of epileptic seizures using \ac{EEG} and \ac{iEEG} signals.

\subsection{Parallel \acp{CNN}}
Parallel \acp{CNN} are composed of one or many convolutional layers, which are executed in parallel and have been previously used in many applications. For example, in the ResNeXt~\cite{Xie2017} family of architectures, parallel blocks containing convolutional layers were used to increase network width, which can decrease the time required to train a \ac{CNN}~\cite{Zagoruyko2016}. When performing multi-modal \ac{DL}, parallel convolutional layers can be used to process different inputs in parallel~\cite{Ngiam2011}, in order to improve network throughput. Specifically for epileptic seizure detection and prediction tasks, parallel convolutional layers have been used to learn high-level representations simultaneously~\cite{Wang2021}. By parallelizing convolutional operations, inference time is greatly reduced compared to current SOTA architecture that rely on sequential convolution layers, as convolution layers form the bottleneck of \ac{CNN} inference.

\subsection{Traditional \ac{EEG}-based Seizure Detection and Prediction Algorithms}
As early as 1996, initial attempts were made to detect seizures using \ac{EEG} signals and traditional \ac{ML} approaches. Using a combination of \acp{ANN} and wavelet transforms, sensitivity values of 76\%~\cite{Webber1996} and 97\%~\cite{Pradhan1996}, were reported using standardized datasets. In the late 2000s and early 2010s, \acp{SVM} encountered growing interest. Namely, when using \acp{SVM} in combination with feature extraction methods such as high-order spectra analysis, wavelet transforms, \acp{FFT}, wavelet decomposition and least-squares parameter estimators~\cite{Chan2008, Netoff2009, Chua2009, Sorensen2010, Chisci2010, Petersen2011, Temko2011, Acharya2011, Kharbouch2011, Liu2012}, promising sensitivity, specificity, and accuracy values $\geq$98.5\% were achieved. More recently, advances in the \ac{DL} domain using \acp{CNN} and \acp{RNN}, have further benefited seizure detection algorithms. Current \ac{SOTA} models are capable of achieving accuracy ranging from 95-100\%~\cite{Ullah2018,Zhao2020,Abiyev2020,Boonyakitanont2019} across multiple datasets.

Early efforts for seizure prediction started in 1970s, where seizure warning systems were designed with logic circuitry to classify extracted features from a series of filters and analog circuitry~\cite{Liss1973, Viglione1973}. To varying degrees of success, a variety of methods have been proposed, including a rule-based method using univariate measures~\cite{Aarabi2012}, spike rate analysis~\cite{Li2013}, positive zero-crossing intervals analysis~\cite{Zandi2013}, statistical dispersion measures~\cite{Bedeeuzzaman2014}, multidimensional probability evolution~\cite{McSharry2002}, circadian concepts via probabilistic forecasting~\cite{Schelter2011}, and a combination of reinforcement learning, online monitoring and adaptive control theory~\cite{Wang2010}. Similarly to seizure detection, many \ac{DL} techniques have also been applied. Notable contributions include the combination of \acp{CNN} and \acp{RNN}, capable of achieving 99.6\% accuracy and a \ac{FPR} of 0.004 per hour~\cite{Daoud2019}. Moreover, supervised deep convolutional autoencoder and bidirectional long short-term memory networks have been used to achieve accuracy, sensitivity, specificity, and precision values between 98-99\%, with F1-values $\geq$0.98. More recently, augmented \ac{DL} network architectures have been used to reduce computational complexity for operation in resource-constrained environments. One such approach, which employs \acp{CNN} with minimizing channels, is capable of achieving 99.47\% accuracy, 97.83\% sensitivity, 92.36\% specificity, with a \ac{FPR} of 0.0764~\cite{Jana2021}. Finally, Siamese models have been used to achieve 88-91\% accuracy on the CHB-MIT dataset~\cite{Dissanayake2020}. We refer the reader to~\cite{Alotaiby2014} for a comprehensive survey of \ac{EEG} seizure detection and prediction algorithms.

\subsection{Hardware Implementations of \ac{EEG}-based Seizure Detection and Prediction Algorithms}
\label{litreview}
Many hardware implementations of epileptic seizure detection and prediction algorithms have been reported using a variety of technologies; namely \ac{FPGA}, \ac{CMOS} and \ac{VLSI}~\cite{Azghadi2020,Kassiri2017}. Complementing traditional hardware implementations, \ac{IMC} architectures, which use memristive crossbar arrays to perform repetitive operations in-memory, have gained increasing popularity in recent years.
Kudithipudi \textit{et al.} implemented a neuromemristive reservoir computing architecture to achieve 90\% accuracy and Merkel \textit{et al.} achieved 85\% accuracy~\cite{Kudithipudi2016,Merkel2014}. Nature-inspired memristive \ac{CA} was implemented by Karamani \textit{et al.} to emulate epilepsy-related phenomena in the brain~\cite{Karamani2018}.

Recent works by Liu \textit{et al.} implemented \ac{FIR} filter bank on memristive crossbar array to achieve 93.46\% seizure detection accuracy and obtained 95\% accuracy by using a memristive crossbar based signal-processing stage combined with linear discriminant classifier~\cite{Liu2020}. Lammie \textit{et al.} pioneered the implementation of \acp{CNN} for seizure prediction using memristor arrays, achieving 77.4\% sensitivity and 0.85 \ac{AUROC} on the CHB-MIT dataset~\cite{Lammie2021}.

Seizure is a chronic, recurring condition that can mostly be prevented through medication before onset \cite{Usman2017}, but even with the best medications, 30\% of the patients are drug-resistant \cite{Fujiwara2015}. Closed-loop brain stimulation has been found to mitigate and even improve symptoms \cite{Zanghieri2021,Heck2014}, but unpredictability of seizure requires a closed-loop prediction system to provide accurate warning with adequate preparation time for stimulation \cite{Nasseri2021}. This calls for the need for fast, low-latency computations, as the changes within the patients can be noticed early-on, in order to start treatments early to improve safety and quality of life \cite{Yang2018}. In doing so, symptoms and subsequent effects can be minimized, including anxiety and social exposition \cite{Pinto2021}. The major limiting factor of seizure detection and prediction algorithms is the reliance on patient specific features, leading to undesirable results when generalized to other patients in the real world \cite{Stirling2021}. With energy efficient computations, it enables the deployment of such systems within wearable devices, so that it can be coupled with the stimulation system, as well as allowing data for a patient to be collected in the long-term to further improve model’s predictions by fine tuning the model to better recognize patient-specific signatures \cite{Peng2021}.

It is known that convolutional layers are the bottlenecks of \acp{CNN}. According to Cong et al., convolutions make up more than 90\% of \ac{CNN} inference~\cite{Cong2014}. Therefore, accelerating convolution is pivotal to efficient \acp{CNN} for future seizure detection/prediction systems. Note that all existing hardware implementations of \ac{CNN} memristive accelerators focus on sequential \acp{CNN}. Memristive crossbar acceleration of parallelized convolution layers and blocks, found in many CNN architectures such as ResNeXt~\cite{Xie2017}, are explored in this work to further reduce inference latency.

\section{Seizure Detection and Prediction System}\label{sec:novel_system}
In this section, we present our seizure detection and prediction system. As shown in Fig.~\ref{workflowl}, our system comprises of five stages, depicted using Fig.~\ref{workflowl}(a)-(e). As the same network architecture, depicted in Fig.~\ref{workflowl}(f), is used for both detection and prediction, and networks are bench-marked using multiple datasets, our proposed system can be reconfigured for both epileptic seizure and prediction tasks. While we briefly detail and discuss signal acquisition and pre- and post-processing stages, here-on-in, the scope of this paper will be largely confined to the accelerator step described in Fig.~\ref{workflowl}(d). We leave a detailed hardware description and evaluation of other stages to future work.

\subsection{Parallel Convolutional Neural Network Architecture}
The primary constraint put on our design was a fixed modular tile size of 64$\times$64. Practically, passive memristor-based analog crossbar tiles of sizes up to 128$\times$64 have been used to perform \acp{VMM}~\cite{Li2018}, however such designs have only been demonstrated using pseudo-crossbars having micron-size electrodes. Such limitations in the maximum viable size are a serious computational scalability challenge with electrodes in the tenth of nanometer range that would prevent sinking large currents through them~\cite{Amirsoleimani2020}. Recently, a 4K memristor analog-grade passive crossbar circuit has been fabricated~\cite{Kim2021}, which comprises several modular 64 x 64 passive crossbar tiles  with ~99\% functional nonvolatile metal-oxide memristors. From an original exploratory investigation, it was determined that for the \ac{RRAM} device being modelled, the largest feasible modular tile size which is able to be programmed using a write-verify scheme was 64$\times$64. Consequently, this fixed modular tile size was used in our designs to minimize the power and area overhead of peripheral circuits and tile interconnects, which are much larger when smaller fixed modular tiles are used.

\subsection{Model Search and Selection}
\label{sec:modelselection}
Most current state-of-the-art \acp{CNN} employ sequential convolution layers, whereby subsequent convolution operations are dependent on results from previous layers. However, in parallel \acp{CNN}, convolution layers can be processed simultaneously, enabling the use of multiple crossbars at the same time. In addition, parallel convolution layers with different kernel sizes enable the network to extract features of varying receptive fields, providing the fully connected layers a diverse and yet compact representation of the features for classification; enabling a reduction in network parameters required. 
\begin{algorithm}[!t]
    \caption{Model Search and Selection Methodology}
	\begin{algorithmic}
	    \renewcommand{\algorithmicrequire}{\textbf{Input:}}
		\renewcommand{\algorithmicensure}{\textbf{Output:}}
	   \Require Fixed modular crossbar tile size ($m \times n$), $\textsc{Obj}_{max}$, objectives to minimize, $\textsc{Obj}_{min}$, additional hardware design constraints, $\mathbf{w}$.
	   \Ensure Optimized network architecture $(L, D, \boldsymbol\alpha, \boldsymbol\beta)$, where $L$ is the number of convolutional layer blocks, $D$ is the number of fully connected layers, $\boldsymbol\alpha$ is a vector containing the sizes of the first kernel for each convolutional layer when parallel convolutional layer execution is performed, and $\boldsymbol\beta$ is a vector containing the number of output neurons for each fully connected layer
		\State \textbf{minimize} $\textsc{Obj}(m, n, L, D, \boldsymbol\alpha, \boldsymbol\beta)$ subject to \textbf{w}.
		\vspace{1ex}
		\Procedure{Network$\_$Architecture}{$m, n, L, D, \boldsymbol\alpha, \boldsymbol\beta$}
		\For{$l = 0$ to $L - 1$} \Comment{For each convolutional layer}
		    \State $\mathbf{C_{in}}_l$ = $m$ \Comment{Input channels}
		    \State $\mathbf{C_{out}}_l$ = floor($n$ / 2) \Comment{Output channels}
		    \If{parallel convolutional layer execution}
		        \State $\mathbf{k}_{l0} = \boldsymbol\alpha_l$, $\mathbf{k}_{l1} = m - 2 - \boldsymbol\alpha_l$ \Comment{Set kernel sizes}
    		\Else
    		    \State $\mathbf{k}_l = m - 1$ \Comment{Set kernel size}
    		\EndIf
		\EndFor
		\For{$d = 0$ to $D - 2$} \Comment{For each fully connected layer}
		    \State $\mathbf{m}_d = \boldsymbol\beta_l$ \Comment{Set number of output neurons}
		\EndFor
		\State $\mathbf{m}_{D - 1} = 2$ \Comment{Last layer}
		\EndProcedure
		\vspace{1ex}
		\Function{Obj}{$m, n, L, D, \boldsymbol\alpha, \boldsymbol\beta, \mathbf{w}$}
		    \State \textbf{maximize} \textsc{Eval}(Net) and \textbf{minimize} \textsc{Params}(Net), \Comment{i.e., determine $L$, $D$, $\boldsymbol\alpha$, and $\boldsymbol\beta$, where \textsc{Eval} determines the validation accuracy, and \textsc{Params} determines the total number of network parameters}
		    \State where,
		    \State Net = \textsc{Network}\_\textsc{Architecture}($m, n, L, D, \boldsymbol\alpha, \boldsymbol\beta$)
		    \Return $\textsc{Obj}_{min}$(Net)
		\EndFunction
	\end{algorithmic}\label{alg:prog_train_alg}
\end{algorithm}

As shown in Fig.~\ref{workflowl}, our proposed \ac{CNN} architecture consists of two parallel convolution kernels. Algorithm~\ref{alg:prog_train_alg} formalizes the methodology used to search for and select the employed model. For our selected model, latency was minimized using $\textsc{Obj}_{min}$. $L$, $D$, and $\boldsymbol\beta$ were fixed to values determined empirically using a preliminary exploratory analysis, and $\boldsymbol\alpha$ was optimized as per Algorithm 1. The following additional hardware design constraints were imposed for our design:  all convolutional layers must be capable of fitting onto one modular crossbar tile, and the total number of required modular crossbar tiles must not exceed 8.

As the convolution operation bottlenecks \ac{CNN} inference, the size of kernels used in parallel convolution layers need to be carefully considered to optimize both network performance and latency. In our proposed architecture, shown in Fig.~\ref{workflowl}(f), we have two parallel convolution layers and one average pooling layer, comprising one convolutional block. To parallelize the two convolution layers, it would be necessary to map the weights of the two convolution layers onto two separate crossbars. As a design choice, we wanted to retain the flexibility of mapping both convolution layers onto the same crossbar, if space complexity is prioritized over latency. Therefore, during the kernel size search, we imposed a constraint of 62, i.e., $m - 2$, for the sum of convolution kernels, as 2 additional rows are designated for implementing the bias for both parallel convolution layers.

When denoting the kernel size of the first parallel convolutional layer as $\alpha$, the kernel size of the second parallel convolutional layer can be expressed as $62 - \alpha$. To determine the optimal network architecture, the University of Bonn’s \ac{EEG} seizure dataset~\cite{Andrzejak2001} was used. Specifically, a 80:20 train validation split was employed, and \textsc{Eval}(Net) was used to determine the 5-fold cross validation accuracy. Seed values of 32 and 8 were arbitrarily set for the network architecture search, to ensure reproducibility of results, and to reduce bias between search and validation.

Empirically, $L=1$, $D=2$, and $\boldsymbol\beta$=[8,] achieved substantial performance. For the single convolutional block, $\boldsymbol\alpha_0$ was varied between 31 and 60. A validation accuracy of 100\% was achieved for all values of $\boldsymbol\alpha_0$, except for $\boldsymbol\alpha_0=60$, which achieved an optimal validation accuracy of 99.375\%. This is not surprising, as the window size of input data is only 64. Therefore, convolution kernel sizes of 60 and 2 provides two extreme and dramatically different receptive fields. In particular, a kernel size of 2, which corresponds to around 10ms of data at 173.61Hz, is likely insufficient to capture local correlation and learn seizure characteristics. The final model was chosen using Occam's razor principle, whereby the simplest model is the best model. Consequently, a kernel size of 32 was selected, as a kernel size 31 would be the simplest to implement due to symmetric convolution kernel sizes; however 32 provides a more diverse receptive field. To further demonstrate the advantage of varied kernel size, a 5-fold cross-validation was performed using a) 64 filters of kernel size 31 b) two parallel convolution layers each with 32 filters of kernel size 30 and 32 (see Fig. \ref{workflowl}). It was observed that both networks are capable of achieving accuracy varying between 99.61\% to 99.83\%, but varied kernel size leads to +0.03\%, -0.01\%, +0.02\% change in performance on Bonn, SWEC-ETHZ and CHBMIT datasets, respectively, compared to using 64 filters of kernel size 31. Although a small degradation in performance is observed for SWEC-ETHZ dataset, improvements are observed for both Bonn and CHBMIT dataset. A net improvement is observed for both seizure detection and prediction using a varied kernel size, while both experiments employ an identical number of weights.

\begin{figure}[!t]
    \centering
    \includegraphics[width=0.5\textwidth]{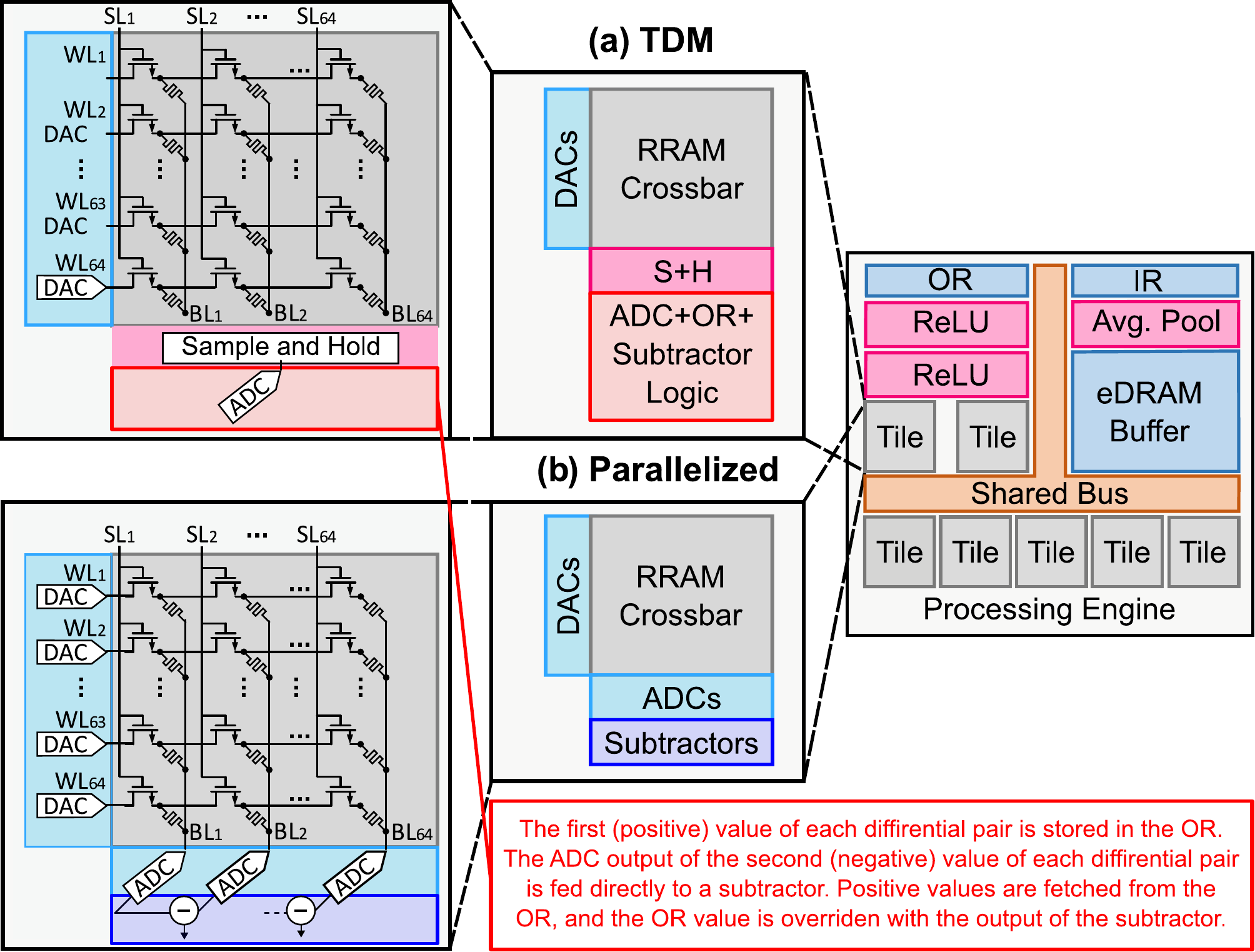}
    \caption{Architecture hierarchy of our memristive DL accelerator with (a) TDM and (b) Parallelized Implementation.}
    \label{fig:block_diagram}
\end{figure}

\subsection{Hardware Architecture Hierarchy}
In Fig~\ref{fig:block_diagram}, we present our hardware architecture hierarchy. The processing engines comprises 7 memristive crossbar array tiles, as well as I/O registers, eDRAM buffers, and peripheral circuits for ReLU, subtract, and average pooling. We present two configurations for our tile, \ac{TDM}, and parallelized. In the \ac{TDM} case, each tile contained a S+H and an \ac{ADC} for reading out column currents, and one \ac{DAC} per row for reading inputs in parallel, as shown in Fig. \ref{fig:block_diagram}(a). In the parallelized case, each tile contains 64 \acp{ADC}, as shown in Fig. \ref{fig:block_diagram}(b).

\section{Software Methodology}\label{sec:software_methodology}
To train and evaluate our epileptic seizure detection and prediction system, we benchmarked our system using one epileptic seizure detection task and two epileptic seizure prediction tasks. For epileptic seizure detection, the University of Bonn’s \ac{EEG} seizure dataset~\cite{Andrzejak2001} was used. For epileptic seizure prediction, the CHB-MIT Scalp \ac{EEG}~\cite{Goldberger2021}, and the long-term SWEC-ETHZ \ac{iEEG}~\cite{Burrello2019} datasets were used.

To perform epileptic seizure detection and prediction, \ac{EEG} and \ac{iEEG} samples can be categorized as either ictal, interictal or preictal. Ictal samples indicate the presence of a seizure, interictal samples are periods between seizures, and preictal samples can be used to detect the onset of a seizure. For epileptic seizure detection, binary classification is performed between ictal and interictal samples. For epileptic seizure prediction, binary classification is performed between preictal and interictal samples. For both epileptic seizure detection and prediction tasks, on account of unbalanced classes, 5-fold cross validation was used to train and validate our network architecture.

\subsection{Training and Evaluation Methodologies}
\subsubsection{Epileptic Seizure Detection}
\label{sec:seizuredetec}
The University of Bonn's \ac{EEG} seizure dataset is comprised of 5 sets (A-E), where set A is normal with open eyes, set B is normal with closed eyes, set C and D is seizure free intervals, and set E is seizure only activities. Each set contains 100 single-channel \ac{EEG} time series of 23.6 seconds, with 4,096 samples in each time series. All data were collected at 173.61 Hz, at a resolution of 12 bits. To perform binary classification between ictal and interictal samples, all samples from sets A and E were used.

Both sets (A and E) were divided into samples of 64 seconds periods and randomly shuffled. No augmentation and pre-processing techniques, such as normalization, were performed, as \acp{CNN} are capable of automatic feature extraction from time-series data and are robust to noise. The lack of need for pre-processing steps implies reduced hardware complexity to perform such operations. Using the network model (with optimal kernel sizes determined in Section \ref{sec:modelselection}), a 5-fold cross-validation strategy was used to determine network's performance. 
To determine performance, the mean of left out set accuracy, sensitivity, specificity, false-positive rate and the \ac{AUROC} across folds of 5-fold cross-validation were reported.

\subsubsection{Epileptic Seizure Prediction}
\label{sec:seizurepred}
The CHB-MIT Scalp \ac{EEG}, and the long-term SWEC-ETHZ \ac{iEEG} datasets were used. The CHB-MIT Scalp \ac{EEG} dataset comprises of 23 cases, which were collected from 22 subjects (5 males, ages 3–22; and 17 females, ages 1.5–19). The last case was obtained 1.5 years after the first, from one of the female subjects~\cite{Goldberger2021}. All signals were sampled at 256Hz with 16-bit resolution, using 23-26 electrodes. During data acquisition, no augmentation steps were performed.

\begin{table}[!t]
\centering
\caption{Overview of cases used to perform epileptic seizure prediction from the CHB-MIT Scalp \ac{EEG} (CHB-MIT) and the long-term SWEC-ETHZ \ac{iEEG} (SWEC-ETHZ) datasets.}
\resizebox{0.5\textwidth}{!}{
\begin{tabular}{lrrrrrr}
\toprule \toprule
\textbf{Patient} & \textbf{Seizures} & \textbf{Interictal Hrs.$^*$} & \textbf{Preictal Hrs.$^*$} & \textbf{Interical Smp.$^\dagger$} & \textbf{Preictal Smp.$^\diamond$} & \textbf{Synthetic Preictal Smp.$^\diamond$} \\
\midrule
\multicolumn{7}{c}{\textbf{CHB-MIT}} \\
\midrule
1 & 7 & 33.74 & 0.43 & 1,898 & 24 & 42 \\
2 & 3 & 32.85 & 0.14 & 1,848 & 8  & 14 \\
3 & 7 & 30.86 & 0.39 & 1,736 & 22 & 37 \\
5 & 5 & 33.85 & 0.30 & 1,904 & 17 & 30 \\
8 & 5 & 14.93 & 0.36 & 840  & 20 & 3 \\
\midrule
\multicolumn{7}{c}{\textbf{SWEC-ETHZ}} \\
\midrule
1 & 2 & 19.91 & 1.00 & 1,120 & 56 & 108 \\
2 & 2 & 19.91 & 1.00 & 1,129 & 56 & 108 \\
3 & 4 & 29.87 & 1.99 & 1,680 & 112 & 216 \\
5 & 4 & 29.87 & 1.99 & 1,680 & 112 & 216 \\
6 & 8 & 69.69 & 3.48 & 3,920 & 196 & 430 \\
\bottomrule \bottomrule
\end{tabular}
}
\begin{tablenotes}
      \item $^*$Hours. $^\dagger$Samples.
\end{tablenotes}
\label{tab:prediction_overview}
\end{table}

\begin{figure}[!t]
     \centering
  \includegraphics[width=0.45\textwidth]{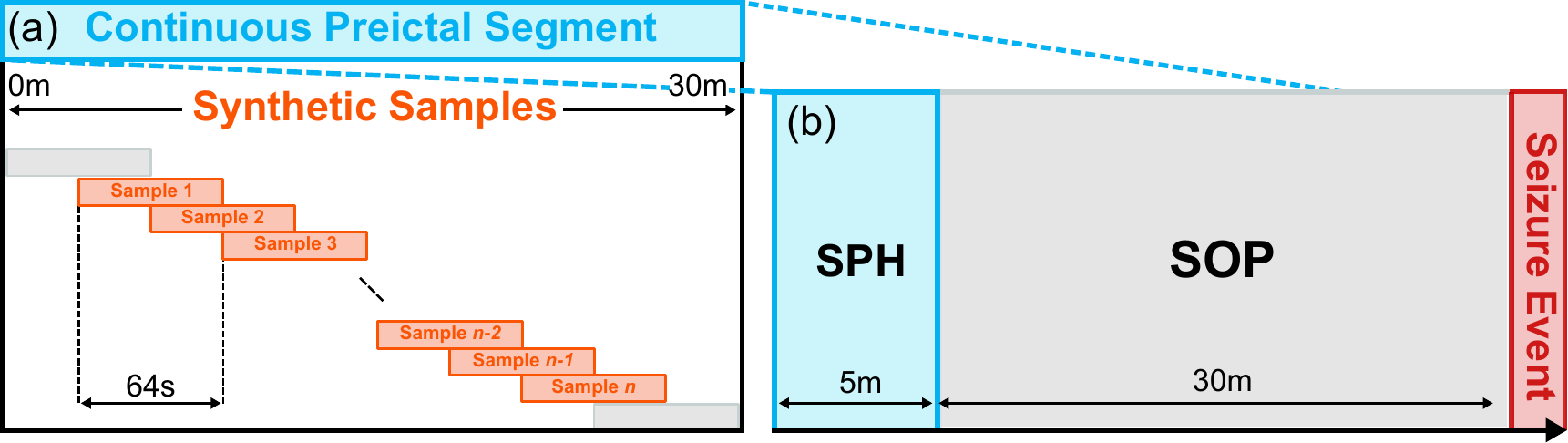}
  \caption{Depiction of (a) our adopted overlapped sampling technique extracting $n$ samples from a continuous preictal segment, and (b) the \ac{SPH} and \ac{SOP} terms. As can be seen, continuous preictal segments are extracted during the \ac{SPH}. All preictal samples that occur during the \ac{SOP} period are discarded.}
  \label{fig:pre_prediction}
\end{figure}

The long-term SWEC-ETHZ \ac{iEEG} dataset comprises of 18 patients with pharmaco-resistant epilepsy, who were evaluated for surgery at the \ac{SWEC} of the University Department of Neurology at the Inselspital Bern~\cite{Burrello2019}. All signals were sampled at either 512Hz or 1025Hz with 16-bit resolution, using 26-100 electrodes. During data acquisition, after analog-to-digital conversion, a digital band-pass filter was used to filter signals between 0.5 and 150Hz using a fourth-order Butterworth filter. Moreover, forward and backward filtering was applied to minimize phase distortion.

Due to computation burden of crossbar simulation, we report the performance using the first 5 viable cases of the the CHB-MIT Scalp \ac{EEG} and long-term SWEC-ETHZ \ac{iEEG} datasets, reducing the computation required, similar to \cite{Truong2018, Wang2021}. In Table~\ref{tab:prediction_overview}, we present an overview of all cases used to perform binary classification between preictal and interictal samples. A case was categorized as viable if it contained valid labels (namely time-stamps) and data files (i.e., no recording files were missing or corrupt). For both datasets, the first 22 channels of each patient were extracted and used. All signals were down-sampled to 256Hz, and a window size (batch size) of 64s was used when extracting samples. After discarding seizures that occur in the first 20-minute monitoring period, a \acf{SOP} of 30m and a \acf{SPH} of 5m were used to extract and label preictal samples for all cases; both of which have previously demonstrated significant performance~\cite{Truong2018}. These terms are defined visually in Fig.~\ref{fig:pre_prediction}. Interictal samples were extracted from one hour recording segments containing no seizures (ictal samples) to reduce class inbalance during training.

Next, 176 features per sample were extracted (8 per channel per window/batch interval): the mean, variance, skewness, kurtosis, coefficient of variation, median absolute deviation of \ac{EEG} amplitude and \ac{RMSA}, and the shannon entropy. Since the input size of the proposed network is 64, the dimensionality of the input data needed to be reduced. A correlation analysis was first performed across the 176 extracted features, but no particular channel could be removed as no strongly correlated channels were discovered. Using \ac{PCA}, linear dimensionality reduction via \ac{SVD} enabled the projection of data to lower dimensional space of 64 principal axes. During training, synthetic preictal samples were generated using an overlapped sampling technique inspired by~\cite{Alotaiby2014}, by sliding a 64s window with a stride of 32s across continuous preictal segments extracted during the \ac{SPH} period, as depicted in Fig.~\ref{fig:pre_prediction}. The same cross-validation training and evaluation strategy and metrics as described in Section \ref{sec:seizuredetec} was employed.

\section{Hardware Methodology}\label{sec:hardware_simulation_methodology}
In this section, we discuss our device technology selection, memristor crossbar array implementations of \acp{CNN}, and present our adopted hardware simulation methodology.

\subsection{Device Technology Selection}
Computing with charge-based computing devices is attractive due to their technological maturity, even though they have a relatively large area footprint even at advanced technology nodes and face severe scaling challenges~\cite{Sebastian2020}. Resistance-based memory, in contrast, can be scaled to the nanometer scale, and has the potential of forming cross-point structures without using access devices, achieving ultra high density. \ac{RRAM} devices are used in our design, as they are widely considered to be the most promising emerging resistance-based memory technology- they operate faster than \ac{PCM}, have a simpler and smaller cell structure than \ac{MRAM} and \ac{CBRAM} devices, and are made of materials that are common in semiconductor manufacturing~\cite{Sebastian2020}.

\subsection{Memristor Crossbar Array Implementations of Parallel CNNs}
Consider the conductance values of a crossbar array as a matrix and input voltages to a crossbar as a vector. The output current from the crossbar, determined using Kirchoff's and Ohm's Law represents the result of the \ac{VMM}. Such operations form the core of \acp{CNN}. Being able to accelerate and parallelize them would facilitate the real-time operation of deeper and heavier neural networks for epileptic seziure detection and prediction in resource-constrained hardware~\cite{RahimiAzghadi2020}.

\begin{figure}[!t]
  \centering
  \includegraphics[width=0.5\textwidth]{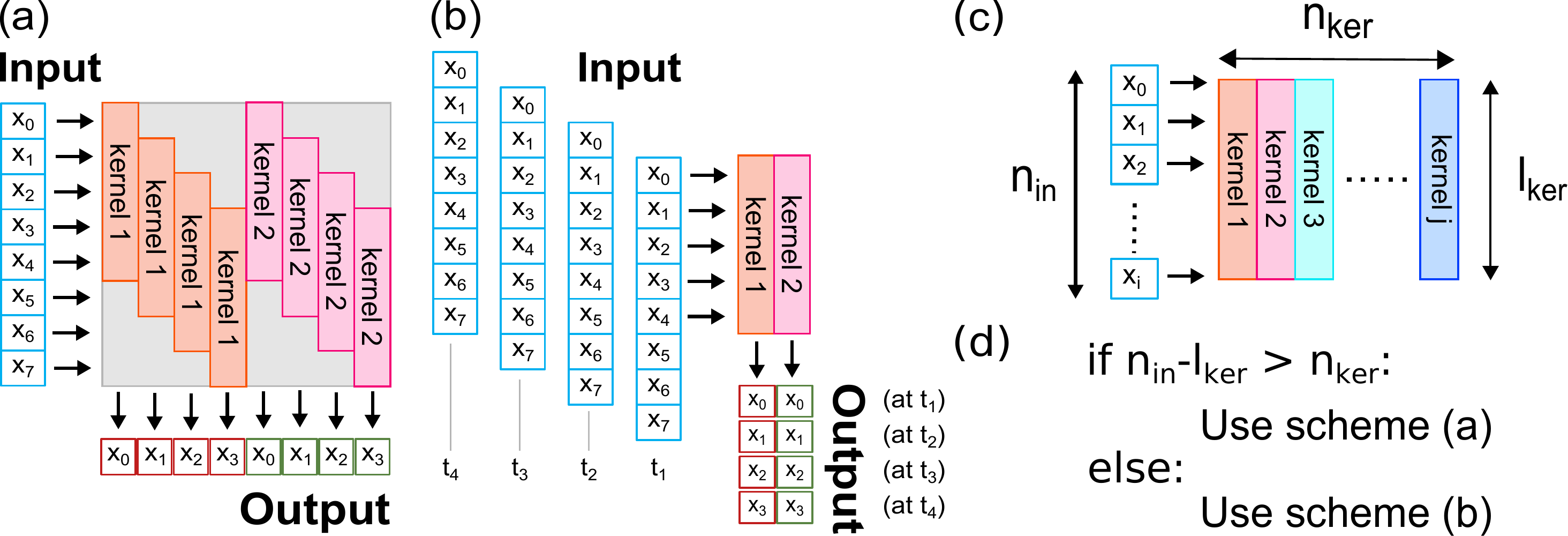}
  \caption{A comparison of possible mapping schemes. (a) visualizes the staggering mapping of convolution weights, which is commonly adopted due to its ability to produce all results within a single pass through the crossbar array. (b) visualizes our proposed mapping scheme, without staggering of convolution weights and sparsity in crossbar, at the cost of increased read/write operations. (c) provides a comparison of methods (a) and (b), visualizing when one method should be chosen over the other.
  }
  \label{mappingcompare}
\end{figure}

\begin{table*}[!t]
\centering
\caption{Crossbar mapping comparison for space and computation trade-off using schemes (a) and (b) in Fig.~\ref{mappingcompare}.}\label{tradeoff_table}
\resizebox{1\textwidth}{!}{%
\begin{tabular}{lrrrrrrrr}
\hline
\toprule \toprule
\multirow{2}{*}{\textbf{Layer}} & \multicolumn{4}{c}{\textbf{Number of Memristor Cell Required}}                                                                                                 & \multicolumn{4}{c}{\textbf{Number of Memristor Cell Required Inc. Sparsity}}                                                                              \\ \cmidrule{2-9}
                       & \textbf{Scheme (a)} & \textbf{Scheme (b)} & \textbf{Area Reduction} & \textbf{Computation Increase} & \textbf{Scheme (a)} & \textbf{Scheme (b)} & \textbf{Area Reduction} & \textbf{Computation Increase} \\
\midrule
\texttt{conv1}                  & 69,696      & 2,112       & 33x                                                       & 33x                                                             & 133,184    & 2,112       & 63x                                                       & 33x                                                             \\
\texttt{conv2}                  & 69,440      & 1,984       & 35x                                                       & 35x                                                             & 145,600    & 1,984       & 73x                                                       & 35x                                                             \\
\texttt{fc1}                    & 17,424      & 17,424      & None                                                      & None                                                            & 17,424      & 17,424      & None                                                      & None                                                            \\
\texttt{fc2}                    & 36         & 36         & None                                                      & None                                                            & 36         & 36         & None                                                      & None                                                            \\
\bottomrule \bottomrule
\end{tabular}%
}
\end{table*}

\begin{figure}[!t]
  \centering
  \includegraphics[width=0.5\textwidth]{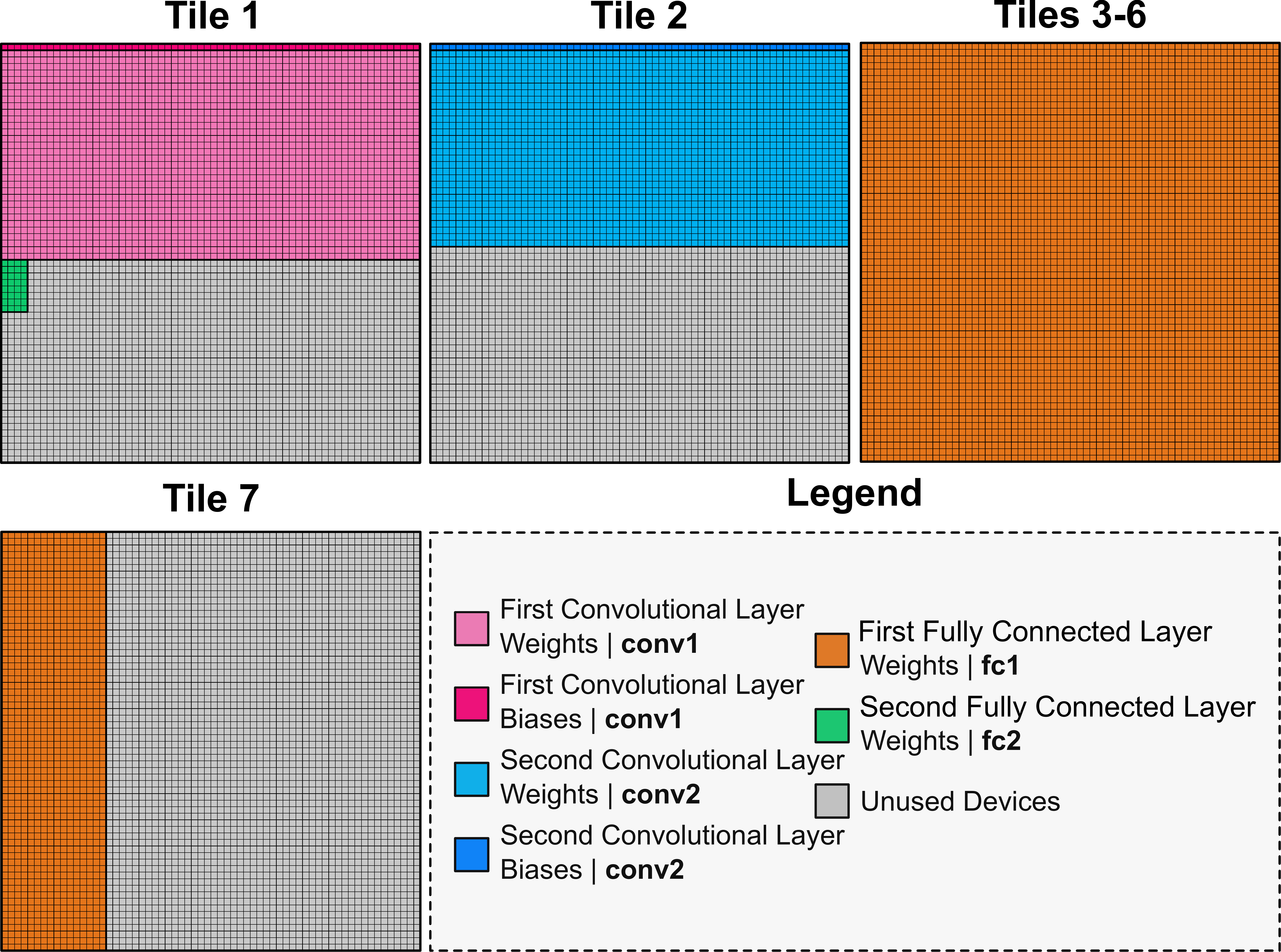}
  \caption{The crossbar parameter mapping layout adopted. Seven 64 $\times$ 64 modular crossbar tiles are utilized. Bias terms of fully connected layers, and the single pooling layer, \texttt{pool1}, are computed using additional digital circuitry. To reduce the number of unused devices, parameters of different layers are shared between tiles.}
  \label{workflow2}
\end{figure}

To represent signed weight matrices on memristive crossbar arrays, as negative conductance values cannot be expressed using analog memristive devices, a differential mapping scheme was adopted, where two columns of memristors are chosen to represent positive and negative weights, respectively. The signed output is thus the arithmetic difference of current from both columns. 
In the case of 1D \acp{CNN}, fully connected and convolutional layers can be decomposed into a series of dot products between inputs, represented as voltages, and weights, represented as memristive conductance.
For convolutional layers, the \texttt{im2col} algorithm~\cite{Chellapilla2006} can be used to map convolutional kernels onto separate crossbar columns. With a single pass, $m$ 1D convolutions can be performed simultaneously, where $m$ represents the number of columns. Average pooling and ReLU operations are performed using additional digital circuitry.

\subsection{Hardware Simulation Methodology}
Based on existing literature from Section~\ref{litreview}, all mapping of convolution kernels onto crossbars are sparse, whereby the convolution kernels form a sparse diagonal matrix, as depicted in Fig.~\ref{mappingcompare}(a). This naive approach is extremely space demanding, as the kernels are staggered multiple times throughout the crossbar array, rendering a lot of memristive cells unused. To reduce the space requirement of mapping scheme (a), one possible approach is to build upon the input-stationary concept. One may remap the crossbar weights during inference and replace them with different kernel weights, while reusing the input fetched from memory.

On the other hand, one may build upon the weight-stationary concept, as depicted in Fig.~\ref{mappingcompare}(b). In this scheme, convolution kernels can be mapped without staggering before inference. For kernels to convolve against different parts of the signal, the input signal slides. The bottleneck of this approach now lies within fetching input data, requiring additional read/write operations on the peripheral of the crossbar compared to mapping scheme (a). The weight-stationary approach is more efficient compared to the input-stationary approach, as crossbar weight writes can be very time and energy consuming, compared to fetching of inputs and staggering them with shifting circuitry. Fig.~\ref{mappingcompare}(c) provides visualization of when one scheme should be adopted over the other.

\begin{table*}[!t]
\centering
\caption{5-Fold cross-validation result for epileptic seizure detection and prediction using our network architecture.}
\resizebox{\textwidth}{!}{%
\begin{tabular}{lrrrrrrrrrrr}
\toprule \toprule
\textbf{Dataset}           & \multicolumn{1}{c}{\textbf{Bonn}}        & \multicolumn{5}{c}{\textbf{CHB-MIT}}                             & \multicolumn{5}{c}{\textbf{SWEC-ETHZ}}                                \\ \cmidrule{2-2} \cmidrule{3-7}  \cmidrule{8-12}
\textbf{Partition}   & \textbf{Set A vs. E} & \textbf{Patient 1} & \textbf{Patient 2} & \textbf{Patient 3} & \textbf{Patient 5} & \textbf{Patient 8} & \textbf{Patient 1} & \textbf{Patient 2} & \textbf{Patient 3} & \textbf{Patient 5} & \textbf{Patient 6} \\ \midrule
Accuracy & 99.84 $\pm$ 0.37 & 99.50 $\pm$ 0.89 & 99.95 $\pm$ 0.11 & 99.95 $\pm$ 0.13 & 99.73 $\pm$ 0.57 & 98.96 $\pm$ 2.33 & 100.00 $\pm$ 0.00 & 100.00 $\pm$ 0.00 & 100.00 $\pm$ 0.00 & 99.86 $\pm$ 0.22 & 100.00 $\pm$ 0.00\\
Sensitivity & 99.87 $\pm$ 0.28 & 98.64 $\pm$ 2.79 & 100.00 $\pm$ 0.00 & 100.00 $\pm$ 0.00 & 99.62 $\pm$ 0.70 & 99.76 $\pm$ 0.54 & 100.00 $\pm$ 0.00 & 100.00 $\pm$ 0.00 & 100.00 $\pm$ 0.00 & 100.00 $\pm$ 0.00 & 100.00 $\pm$ 0.00\\
Specificity & 99.80 $\pm$ 0.45 & 99.73 $\pm$ 0.37 & 100.00 $\pm$ 0.00 & 99.93 $\pm$ 0.15 & 99.77 $\pm$ 0.52 & 97.38 $\pm$ 5.85 & 100.00 $\pm$ 0.00 & 100.00 $\pm$ 0.00 & 100.00 $\pm$ 0.00 & 99.77 $\pm$ 0.39 & 100.00 $\pm$ 0.00\\
FP per Hour & N/A & 0.13 $\pm$ 0.17 & 0.00 $\pm$ 0.00 & 0.03 $\pm$ 0.07 & 0.10 $\pm$ 0.22 & 0.53 $\pm$ 1.19 & 0.00 $\pm$ 0.00 & 0.00 $\pm$ 0.00 & 0.00 $\pm$ 0.00 & 0.08 $\pm$ 0.13 & 0.00 $\pm$ 0.00 \\
AUROC & 99.84 $\pm$ 0.37 & 99.31 $\pm$ 1.06 & 100.00 $\pm$ 0.00 & 99.82 $\pm$ 0.39 & 99.63 $\pm$ 0.79 & 99.04 $\pm$ 2.15 & 100.00 $\pm$ 0.00 & 100.00 $\pm$ 0.00 & 100.00 $\pm$ 0.00 & 99.84 $\pm$ 0.25 & 100.00 $\pm$ 0.00\\
\bottomrule \bottomrule
\end{tabular}%
}
\label{tab:softwaresummary}
\end{table*}

A comparison of the naive approach and our proposed weight-stationary approach is performed for our network architecture in Table~\ref{tradeoff_table}. As can be observed, the number of memristor cells required for scheme (b) (depicted in Fig.~\ref{mappingcompare} (b)) is significantly smaller, due to the compact nature of the mapping. This comes, however, at the cost of 33x increase in computation. When taking sparsity, i.e. unused memristors depicted by the gray background in Fig.~\ref{mappingcompare} (a), into consideration, scheme (b) demonstrates even more significant reduction, i.e. 63x-73x fewer memristors required, while the computation increase remains constant.  Unlike convolutional layers, fully connected layers do not involve sliding of signals, so \acp{VMM} for fully connected layers were implemented using the naive scheme (a). Using scheme (b), we mapped convolutional kernels within our trained network onto crossbars tiles of 64$\times$64. While scheme (b) was chosen for our hardware design, if scheme (a) were chosen with different $n_{ker}$ and $l_{ker}$ values, or the added space complexity is not of concern, the staggered weights of scheme (a) would enable all rows of the crossbars to be employed simultaneously. By choosing the input size of our network to be 64, we maintain the flexibility of mapping with scheme (a) to  make use of all crossbar rows simultaneously.

As Fig.~\ref{workflow2} demonstrates, for parallel convolution layers to be accelerated simultaneously, it was necessary to map the weights of the \texttt{conv1} and \texttt{conv2} onto two separate crossbar tiles. The weight of the \texttt{fc1} layer is a matrix of 1088$\times$8, and using a differential weight scheme, would require 1088$\times$16 memristors. The weight matrix can be further divided into 17 sections of 64$\times$16 weights. To maximize the usage of each 64$\times$64 crossbar array, 4 sections of 64$\times$16 weights can be stacked horizontally onto each crossbar, requiring a total of 5 crossbar tiles.

Since there are unused memristors on the convolution tiles and \texttt{fc2} layer operations are not performed immediately after convolution operations, we decided to map the weights of \texttt{fc2} onto the convolution layer tile, instead of using another tile. Note that since the simulation serves as a validation for proof-of-concept, we decided to use the same dimensions for all 7 crossbar tiles. We do recognize that tile 1, 2 and 7 have many unused memristor devices, as a result, performing small VMMs on a large switch matrix. This leads to large power overhead due to high amortized ADC/DAC power over a small matrix and charge/discharge of long row and column wires without using full length for computation. To address such problem in a real medical device, instead of using square tiles, tile 1, 2 and 7 can be easily mapped onto rectangular tiles of the exact required dimensions.

\subsection{Impact of Device and Crossbar Non-Idealities}
Memristors and memristive crossbar arrays are prone to numerous device and circuit non-idealities which have been demonstrated to severely impact the performance of memristive \ac{DL} accelerators~\cite{Krestinskaya2019}. Consequently, they should be comprehensively simulated prior to circuit-level realization. 
In this paper, preliminary simulations were performed using the MemTorch~\cite{Lammie2020} simulation framework, and comprehensive simulations of the system using passive crossbar arrays were performed using the crossbar array model provided by~\cite{Chen2013}. Non-idealities considered include input and output resolutions, weight write resolution, weight write deviation, stuck $R_{\text{ON}}/R_{\text{OFF}}$ devices, line and source resistance, and conductance range variation. 

Other memristive phenomena, such as the dynamic behavior of switched memristive neural networks after programming~\cite{Cheng2021}, and read disturbance~\cite{Shim2020}, are not accounted for, as practical metal-oxide memristors are endurance-limited, during programming a write-verify scheme is used, and during inference, all \ac{BL} voltages are constrained to have a maximum absolute amplitude of 0.3V~\cite{Shim2020}.

\subsection{Stuck Weight Offsetting Methodology}
Stuck $R_{\text{ON}}/R_{\text{OFF}}$ weights are known to cause significant network performance degradation in memristive crossbar arrays. Existing works have demonstrated performance recovery through a variety of techniques. In 2014, Kannan et al. took inspiration from SRAM/DRAM technologies and repaired crossbar defects using redundant rows and columns \cite{kannan2014detection}. In 2017, Liu et al. proposed to identify significant weights before applying a retraining and remapping algorithm \cite{liu2017rescuing}. In 2018, Xia et al. proposed a mapping algorithm with inner fault tolerance to leverage the differential mapping scheme of crossbar arrays to tolerate faults \cite{xia2017stuck}. In 2019, Zhang et al. proposed the use of matrix transformations to reduce the magnitude of error introduced by stuck-at-fault devices \cite{zhang2019handling}. Also in 2019, Yeo et al. modified conventional transimpedance amplifiers to detect when abnormal current is detected at a particular column due to stuck-at-fault devices and repair by retraining the network with the known defects \cite{yeo2019stuck}. Among those works, significant hardware or software overhead is introduced through rewriting and tuning of weights, retraining of networks or using additional circuitry.

To minimize the overhead, we propose stuck weight offsetting, which improves upon the inner fault tolerance method. Inner fault tolerance first identifies all available (non stuck-at-fault) devices and initializes them to default values. Then, the scheme goes through all available devices and adjusts each value such that the represented values cannot be made any closer to the target matrix parameter. Intuitively, this serves to minimize the incorrect contribution of the $R_{\text{ON}}/R_{\text{OFF}}$ weight. We propose to bypass the initialization of available devices to default values and to focus on the complementary weight of stuck-at-fault devices only. Before writing any weights to the crossbar, all stuck-at-fault devices are identified. For each stuck-at-fault device, if the complementary weight is not stuck-at-fault, we calculate its complementary weight to minimize the difference between represented value and target value. All calculated values, along with normal weights, are then written onto the crossbar. This modification reduces overhead by two means. First, all crossbar weights are only required to be written once, as opposed to twice in the inner fault tolerance method (from default to adjusted). Second, our method focuses on complementary weights for stuck-at-fault devices only, as opposed to all available devices for all target parameters. This method incurs minimum additional computational cost, and does not require retraining. 

\begin{table}[!t]
\centering
\caption{Comparison of our baseline software model against SOTA for Seizure Detection using the University of Bonn dataset}\label{comparison_SOTA_detection}
\resizebox{\columnwidth}{!}{%
\begin{tabular}{lcrcrr}
\toprule \toprule
\textbf{Paper}                                                                                          & \textbf{Pre-processing} & \textbf{Method} & \textbf{Parallelization} & \textbf{Parameters} & \textbf{Accuracy (\%)} \\ 
\midrule
Ullah \textit{et al.} (2018)      & \ding{51}                    & 1D-CNN          &           \ding{55}           & 21,436               & 99.90           \\
We \textit{et al.} (2018)     & \ding{51}                    & 1D-CNN          &         \ding{55}          & 16,778,144            & 92.00           \\
Abdelhameed \textit{et al.} (2018)                                        & \ding{51}                    & 2D-CNN          &         \ding{55}            & 106,388              & 98.00              \\
Liu \textit{et al.} (2019) & \ding{51}                    & 2D-CNN          &          \ding{55}           & N/R$^*$                  & 99.60           \\
Turk \textit{et al.} (2019)               & \ding{51}                    & 2D-CNN          &         \ding{55}              & 1,603,080             & 99.45           \\
Abdelhameed \textit{et al.} (2021)                                             & \ding{51}                    & 2D-CNN          &       \ding{55}         & 10,304,467            & \textbf{100.00}    \\
\midrule
Ours                                                                                                    &           \ding{55}  & 1D-CNN          & \ding{51}             & \textbf{10,778}      & 99.84           \\ 
\bottomrule \bottomrule
\end{tabular}
}
\begin{tablenotes}
      \item $^*$Not reported.
\end{tablenotes}
\end{table}

\begin{table*}[!t]
\renewcommand{\arraystretch}{1}
\centering
\caption{Comparison against SOTA for Seizure Prediction using the SWEC-ETHZ and CHB-MIT datasets}\label{comparison_SOTA_prediction}
\begin{tabular}{lccrrrrr}
\toprule \toprule
\textbf{Paper}                                                                                           & \textbf{Method} & \textbf{Parallelized} & \textbf{Parameters} & \textbf{Sensitivity (\%)} & \textbf{Specificity (\%)} & \textbf{Accuracy (\%)} & \textbf{FPR$^\dagger$}\\ 
\midrule
\multicolumn{8}{c}{\textbf{CHB-MIT}} \\
\midrule
\noindent \cite{Truong2018} & 2D-CNN & \ding{55} & N/R$^\diamond$ & 81.20 & N/R$^\diamond$ & N/R$^\diamond$ & \textbf{0.16} \\
\cite{Wen2018} * & 2D-CNN & \ding{55} & N/R$^\diamond$ & N/R$^\diamond$ & N/R$^\diamond$ & 92.00 & N/R$^\diamond$ \\
\cite{Hossain2019}          & 2D-CNN & \ding{55}  & 49,560    & 82.71 & 88.21 & 98.19 & N/R$^\diamond$ \\
\cite{Cao2019} *            & 2D-CNN & \ding{55}  & N/R$^\diamond$       & 88.80  & 88.60  & 88.70  & N/R$^\diamond$ \\
\cite{Tian2019} *           & 3D-CNN & \ding{55}  & 28,459,615 & 96.66 & 99.14 & 98.33 & N/R$^\diamond$ \\
\cite{Liang2020} *          & 2D-CNN & \ding{55}  & 9,695,012  & 84.00    & 99.00    & 99.00 & 0.2    \\
\cite{Wang2021}             & 1D-CNN & \ding{51} & 105,538   & 95.55 & \textbf{99.68} & \textbf{99.64} & N/R$^\diamond$ \\ \midrule
Ours                           & 1D-CNN & \ding{51} & \textbf{10,778}    & \textbf{99.24} & 98.68 & 99.01 & 0.47 \\
\midrule
\multicolumn{8}{c}{\textbf{SWEC-ETHZ}}\\
\midrule
\cite{Burrello2020}
* & Ensemble HD & \ding{55}  & N/R$^\diamond$     & 96.38 & 97.31 & 96.85 & N/R$^\diamond$ \\
\cite{Wang2021}     & 1D-CNN      & \ding{51} & 105,538 & 94.57 & \textbf{99.86} & \textbf{99.81} & N/R$^\diamond$ \\ \midrule
Ours                     & 1D-CNN      & \ding{51} & \textbf{10,778}  & \textbf{98.22} & 97.02 & 97.54 & \textbf{0.99} \\

\bottomrule \bottomrule
\end{tabular}
\begin{tablenotes}
      \item *Indicates the results are reported across the entire dataset and patient-wise performance was not reported. $^\dagger$False positive rate (per hour). $^\diamond$Not reported.
\end{tablenotes}
\end{table*}

\begin{figure*}
\centering
\includegraphics[width=0.85\textwidth]{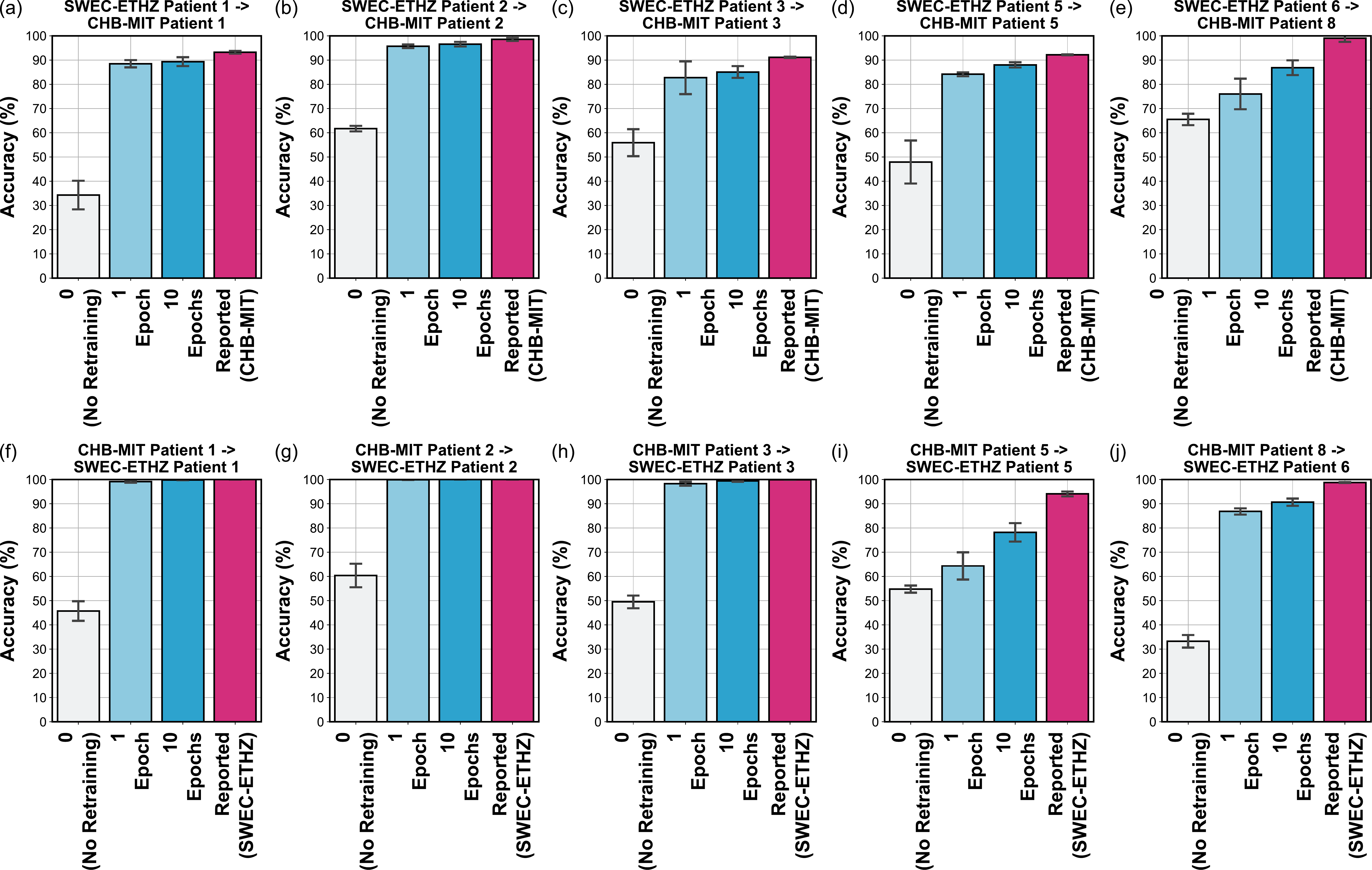}
\caption{The ability of our trained networks to generalize between different datasets when performing epileptic seizure prediction. The cross validation accuracy is reported for networks which have not been retrained, and for networks that have been retrained after 1 and 10 training epochs, respectively, when transfer learning was performed. In addition, the standard evaluation accuracy is reported for each dataset and patient, to facilitate comparisons.}
\label{fig:transfer_learning}
\end{figure*}

\subsection{Quantization Aware Training for Lower Resolution Systems}
A high resolution system is often not feasible to deploy on edge devices, given power consumption constraints and sampling frequency requirements, which are fundamental tradeoffs for resolution in \acp{DAC} and \acp{ADC}. However, lower resolution systems with improved power and frequency performance can exhibit performance degradation. This effect was observed for some patients, and more details can be found in Section \ref{sec:QAT}. For significant performance degradation (a degradation of 5\% or more compared to full resolution system), we propose to perform \acf{QAT} prior to mapping the weights onto memristive crossbar arrays~\cite{Brevitas}.
During \ac{QAT}, we quantized the convolutional and fully connected layers of the network to the resolution equivalent to or even lower than that of the resolution of the crossbar weights and \ac{ADC}/\ac{DAC} resolution. Quantized layers are implemented using the Brevitas library~\cite{Brevitas}, which provides PyTorch-compatible convolution and fully connected layers of specified weight resolutions. In addition, inputs to the network were quantized, while intermediate outputs remained not quantized. Network architecture and other training parameters remained unchanged.

\section{Results and Discussion}\label{sec:results_discussion}
Prior to the investigation of device and crossbar non-idealities, we report baseline software results for epileptic seizure detection and prediction using our network architecture, in Table~\ref{tab:softwaresummary}. 5-fold cross-validation was performed using a different seed to eliminate bias on the first fold. To demonstrate the generalizability of the designed network to different domains and patients, the same architecture was applied for seizure detection and prediction. Unlike the Bonn dataset, both the CHB-MIT and SWEC-ETHZ datasets are multi-channel \ac{EEG} datasets with larger memory and computation requirements within the time domain. In order to reduce the time and memory complexity, pre-processing steps as described in Section \ref{sec:seizurepred} were applied to transform the dataset into frequency domain. The shown results suggest that the proposed network is sufficient and can generalize well for both detection and prediction. 

\subsection{Comparisons Against SOTA Software Implementations}
In Tables~\ref{comparison_SOTA_detection} and \ref{comparison_SOTA_prediction}, we compare our baseline software implementations that use full precision (32-bit) floating-point parameters against other software implementations in literature for epileptic seizure detection and prediction, respectively. 
As shown in the Tables, for epileptic seizure detection we achieve \ac{SOTA} performance in 3/4 criteria, while for prediction we obtain SOTA performance in 3/6 criteria. Specifically, for detection, our network architecture is able to achieve an accuracy of 99.84\% across all samples without any pre-processing steps, while requiring only 10,778 parameters. This is $\sim$2x fewer parameters  than the smallest model in~\cite{Ullah2018}, which achieved a slightly higher accuracy of 99.90\%, while employing various pre-processing steps. Except for the model used in \cite{Abdelhameed2021}, which achieves a 100\% accuracy, but requires over 10M parameters, all the other models  shown in Table~\ref{comparison_SOTA_detection}, achieve lower accuracy values despite significantly higher number of network parameters.

\begin{figure*}[!t]
  \centering
  \includegraphics[width=1\textwidth]{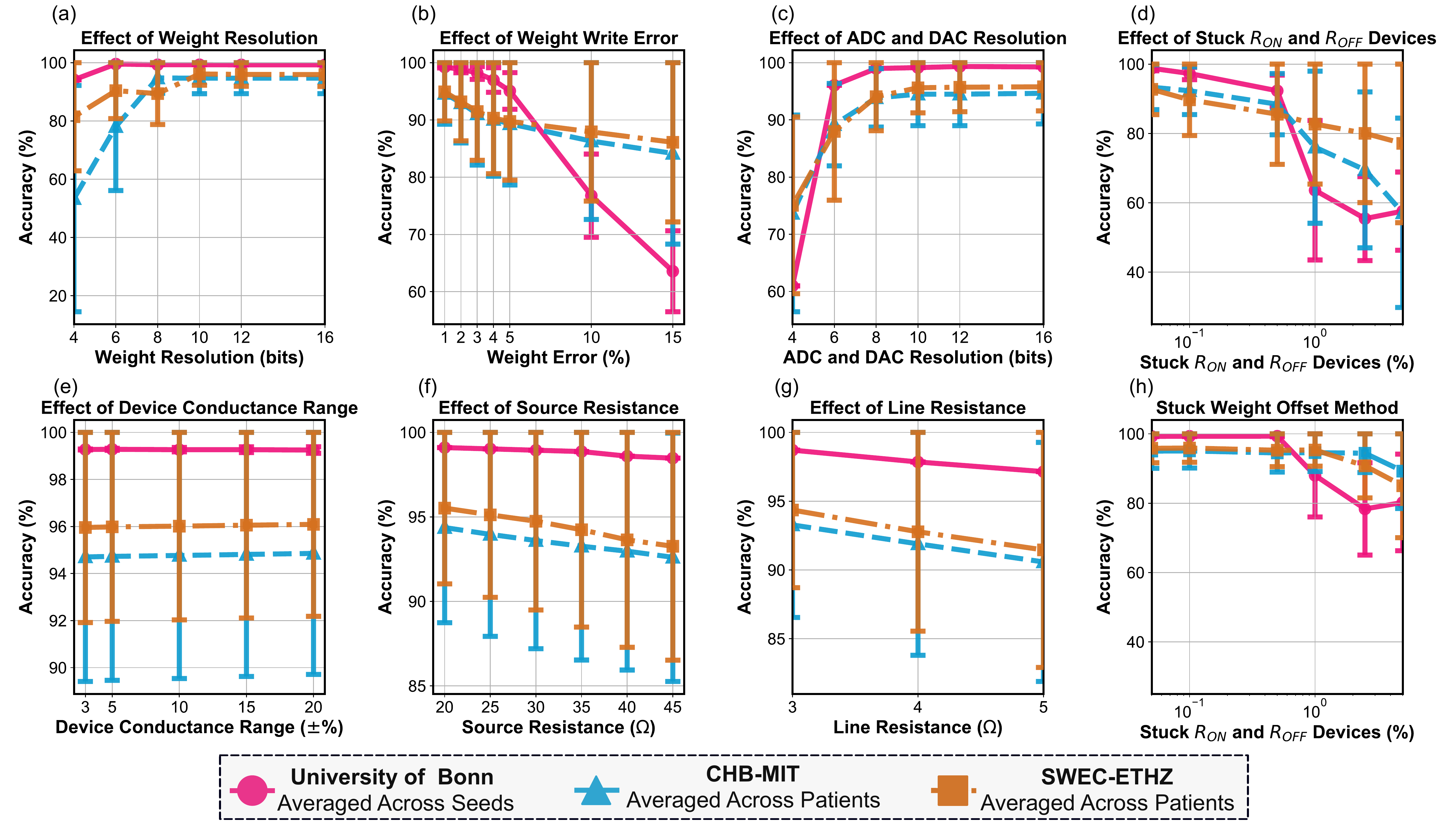}
  \caption{The impact of all (a-g) non-idealities on the University of Bonn, CHB-MIT, and SWEC-ETHZ datasets. (h) summarizes performance recovery by applying our proposed stuck weight offsetting to address the performance degradation of stuck-at fault devices. For the University of Bonn dataset, each data-point shows the mean and standard deviation across five arbitrary seed values: 5, 6, 7, 8, and 9.}
  \label{nonidealities_result}
\end{figure*}

For epileptic seizure prediction, pre-processing is performed. Across both datasets, our network architecture achieves the highest sensitivity while requiring the fewest number of parameters. We report close specificity and accuracy values to~\cite{Wang2021}, which has also used a 1D-CNN architecture with parallelization, but needs $\sim$10x more parameters. Finally, we report the highest \ac{FPR} across both datasets, however, unlike previous works, we performed no post-processing steps, which may cause this. Also, only two out of the nine previous works have reported their \ac{FPR}, which makes the comparison incomplete. 
When mapping trained parameters to ideal crossbars with fully analog devices without any device or circuit non-idealities, the same results were achieved.

\subsection{Generalization Between Datasets}
To determine whether or not our trained networks have the ability to generalize, we evaluated the performance of networks trained using the CHB-MIT dataset on the SWEC-ETHZ dataset, and vice-versa in Fig.~\ref{fig:transfer_learning}. In addition, we report the cross validation accuracy for networks which have been retrained using transfer learning. To perform transfer learning, parameters were frozen for all layers except the last two fully connected layers, and the weights and biases of the last two fully connected layers were re-trained using the training set of the evaluation dataset. Direct evaluations to/from either of these datasets and the University of Bonn dataset were not made, as the University of Bonn dataset is used for epileptic seizure detection and not prediction, and it is structured differently.

\subsection{Quantization-Aware Training}\label{sec:QAT}
\begin{figure*}[!t]
     \centering
    \includegraphics[width=1\textwidth]{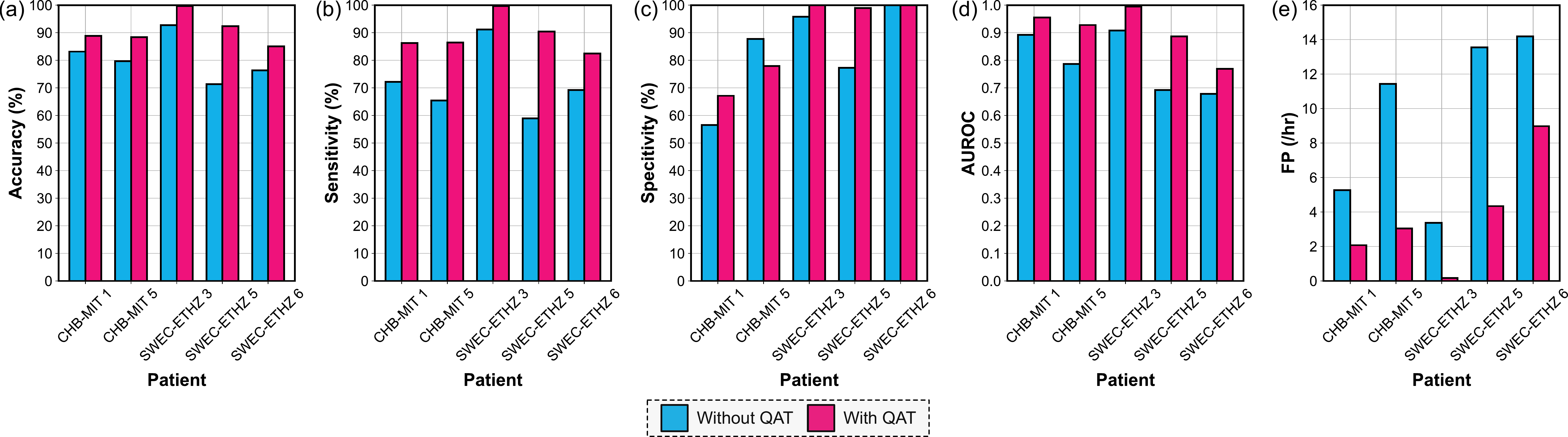}
    \caption{The impact of \ac{QAT} on our network architecture tasked for epileptic seizure prediction (a-e) evaluated using the CHB-MIT and SWEC-ETHZ datasets when network parameters are quantized to 6-bit fixed-point resolution. Only patients that exhibited a degradation of 5\% or more when quantized to 6-bit fixed-point resolution (from full-precision floating-point) were investigated.}
    \label{fig:qat}
\end{figure*}

To demonstrate the effectiveness of \ac{QAT}, we evaluated the performance of our network architecture when trained with and without \ac{QAT}. Comparisons are made in Fig.~\ref{fig:qat}. During \ac {QAT} training, inputs and network weights were reduced to 6-bit resolution, while network architecture and other training parameters were held constant, as described in Fig.~\ref{workflowl}(f).
The accuracy, sensitivity, specificity, AUROC, and \ac{FPR} metrics were all reported and compared. When using 6-bit \acp{ADC} and \acp{DAC}, it can be observed that for all patients and metrics, except for specificity of patient 5 from the CHB-MIT dataset, \ac{QAT} network yields significant performance improvements.

\subsection{Effects of Non-Idealities on System Performance}
Fig.~\ref{nonidealities_result} provides a summary of the impact of non-idealities on our system for epileptic detection and prediction.
For the University of Bonn dataset, as samples between patients are not explicitly distinguished, the mean and standard deviation of test set accuracy is reported across samples using five arbitrarily chosen seed values. For the CHB-MIT and SWEC-ETHZ datasets, the mean and standard deviation of test set accuracy is reported across samples for the first five viable patients of each dataset, respectively. 
Across datasets, some patients were observed to be more robust to non-idealities than others. This was observed in our investigations for patients 1, 2, 3 from the SWEC-ETHZ dataset, and patient 2 from the CHB-MIT dataset, for which non-idealities have minimal impact. For the rest of the patients, however, no clear pattern was established with regards to robustness against non-idealities. We attribute the varying degree of effectiveness between patients to underlying patient specific signatures.

\begin{table*}[!t]
\centering
\caption{Power, area, and latency metrics for the simulated memristive DL accelerator using a 22 nm \ac{CMOS} process. Using our TDM architecture, \acp{VMM} are performed in $\mathcal{O}(n)$, where $n$ is the number of columns of the output vector. Using our parallelized architecture, \acp{VMM} are performed in $\mathcal{O}(1)$.
} \label{table:power_area_latency_requirements}
\resizebox{1\textwidth}{!}{%
\begin{tabular}{l|c|crrrrr|crrrrr}
\toprule \toprule
\multirow{3}{*}{\textbf{Component}} & \multicolumn{1}{l}{\multirow{3}{*}{\textbf{ Params.}}} & \multicolumn{6}{|c}{\textbf{Time-Division Multiplexing (TDM) }} & \multicolumn{6}{|c}{\textbf{Parallelized }} \\
\cmidrule(lr){3-14}
 & \multicolumn{1}{l|}{} & \textbf{Specification} & \begin{tabular}[c]{@{}r@{}}\textbf{Area}\\\textbf{(mm$^2$)}\end{tabular} & \begin{tabular}[c]{@{}r@{}}\textbf{Power}\\\textbf{(mW)}\end{tabular} & \begin{tabular}[c]{@{}r@{}}\textbf{Latency}\\\textbf{(us)*}\end{tabular} & \begin{tabular}[c]{@{}r@{}}\textbf{Total Latency}\\\textbf{(us)}\end{tabular} & \begin{tabular}[c]{@{}r@{}}\textbf{Energy}\\\textbf{(uJ)}\end{tabular} & \textbf{Specification} & \begin{tabular}[c]{@{}r@{}}\textbf{Area}\\\textbf{(mm$^2$)}\end{tabular} & \begin{tabular}[c]{@{}r@{}}\textbf{Power}\\\textbf{(mW)}\end{tabular} & \begin{tabular}[c]{@{}r@{}}\textbf{Latency}\\\textbf{(us)*}\end{tabular} & \begin{tabular}[c]{@{}r@{}}\textbf{Total Latency}\\\textbf{(us)}\end{tabular} & \begin{tabular}[c]{@{}r@{}}\textbf{Energy}\\\textbf{(uJ)}\end{tabular} \\
 \midrule
\multirow{2}{*}{DAC} & Resolution & 6 bits & \multirow{2}{*}{2.58E+01} & \multirow{2}{*}{2.69E+03} & \multirow{2}{*}{8.00E-04} & \multirow{2}{*}{2.15E+00} & \multirow{2}{*}{5.78E+00} & 6 bits & \multirow{2}{*}{2.58E+01} & \multirow{2}{*}{2.69E+03} & \multirow{2}{*}{8.00E-04} & \multirow{2}{*}{3.36E-02} & \multirow{2}{*}{9.03E-02} \\
 & Number & 7x64 &  &  &  &  &  & 7x64 &  &  &  &  &  \\
 \midrule
\multirow{3}{*}{ADC} & Resolution & 6 bits & \multirow{3}{*}{4.62E+00} & \multirow{3}{*}{7.00E+01} & \multirow{3}{*}{1.00E-01} & \multirow{3}{*}{2.69E+02} & \multirow{3}{*}{1.88E+01} & 6 bits & \multirow{3}{*}{2.96E+02} & \multirow{3}{*}{4.48E+03} & \multirow{3}{*}{1.00E-01} & \multirow{3}{*}{6.00E-01} & \multirow{3}{*}{2.69E+00} \\
 & Number & 7 &  &  &  &  &  & 7x64 &  &  &  &  &  \\
 & Frequency & 10MHz &  &  &  &  &  & 10MHz &  &  &  &  &  \\
 \midrule
ReLU & Number & 2 & 9.60E-03 & 3.28E-02 & 9.80E-02 & 9.80E-02 & 3.22E-06 & 2 & 9.60E-03 & 3.28E-02 & 9.80E-02 & 9.80E-02 & 3.22E-06 \\
\midrule
Average Pool & Number & 1 & 3.83E-04 & 1.59E+00 & 8.49E-05 & 8.49E-05 & 1.35E-07 & 1 & 3.83E-04 & 1.59E+00 & 8.49E-05 & 8.49E-05 & 1.35E-07 \\
\midrule
Adder & Number & 10 & 5.34E-03 & 1.74E-02 & 3.06E-04 & 6.13E-04 & 1.06E-08 & 10 & 5.34E-03 & 1.74E-02 & 3.06E-04 & 6.13E-04 & 1.06E-08 \\
\midrule
Subtractor & Number & 7 & 2.46E-04 & 2.87E-01 & 3.34E-04 & 1.28E-01 & 3.69E-05 & 7x32 & 7.88E-03 & 9.20E+00 & 3.34E-04 & 2.01E-03 & 1.85E-05 \\
\midrule
S+H$^\dagger$ & Number & 7x64 & 8.98E-06 & 3.81E-03 & 8.33E-04 & 5.00E-03 & 1.90E-08 & 7x64 & 8.98E-06 & 3.81E-03 & 8.33E-04 & 5.00E-03 & 1.90E-08 \\
\midrule
\multirow{2}{*}{eDRAM Buffer} & Size & 2KB & \multirow{2}{*}{4.72E-03} & \multirow{2}{*}{1.81E+01} & \multirow{2}{*}{1.15E-04} & \multirow{2}{*}{2.30E-04} & \multirow{2}{*}{4.17E-06} & 2KB & \multirow{2}{*}{4.72E-03} & \multirow{2}{*}{1.81E+01} & \multirow{2}{*}{1.15E-04} & \multirow{2}{*}{2.30E-04} & \multirow{2}{*}{4.17E-06} \\ 
 & Bus Width & 128 &  &  &  &  &  & 128 &  &  &  &  &  \\
 \midrule
eDRAM-Tile Bus & Number & 192 & 4.50E-03 & 3.5E+00 & 9.02E-05 & 9.02E-05 & 3.16E-07 & 192 & 4.50E-03 & 3.5E+00 & 9.02E-05 & 9.02E-05 & 3.16E-07 \\
\midrule
IR$^\dagger$ & Size & 1KB & 8.10E-01 & 6.74E-01 & 8.21E-05 & 1.64E-04 & 1.11E-07 & 1KB & 8.10E-01 & 6.74E-01 & 8.21E-05 & 1.64E-04 & 1.11E-07 \\
\midrule
OR$^\dagger$ & Size & 512B & 8.70E-04 & 4.18E-01 & 8.21E-05 & 1.64E-04 & 6.87E-08 & 512B & 8.70E-04 & 4.18E-01 & 8.21E-05 & 1.64E-04 & 6.87E-08 \\
\midrule
\multicolumn{14}{c}{\textbf{Scenario: $\bar{R_{\text{ON}}}$}} \\
\toprule \toprule
\multirow{3}{*}{Crossbar} & Number & 7 & \multirow{3}{*}{2.87E-04} & \multirow{3}{*}{8.67E+00} & \multirow{3}{*}{2.03E-03} & \multirow{3}{*}{5.82E+01} & \multirow{3}{*}{5.06E-01} & 7 & \multirow{3}{*}{2.87E-04} & \multirow{3}{*}{8.69E+00} & \multirow{3}{*}{2.03E-03} & \multirow{3}{*}{1.30E-01} & \multirow{3}{*}{1.13E-03} \\
 & Size & 64x64 &  &  &  &  &  & 64x64 &  &  &  &  &  \\
 & Bits per cell & 32 &  &  &  &  &  & 32 &  &  &  &  &  \\
 \midrule
\textbf{Total} &  &  & \textbf{3.13E+01} & \textbf{2.79E+03} & \multicolumn{2}{r}{\textbf{3.29E+02}} & \textbf{9.19E+02} &  & \textbf{3.22E+02} & \textbf{7.21E+03} & \multicolumn{2}{r}{\textbf{8.70E-01}} & \textbf{6.27E+00} \\
\toprule \toprule
\multicolumn{14}{c}{\textbf{Scenario: $(\bar{R_{\text{ON}}} + \bar{R_{\text{OFF}}}) / 2$}} \\
\toprule \toprule
\multirow{3}{*}{Crossbar} & Number & 7 & \multirow{3}{*}{2.87E-04} & \multirow{3}{*}{4.35E+00} & \multirow{3}{*}{6.07E-03} & \multirow{3}{*}{1.74E+02} & \multirow{3}{*}{7.58E-01} & 7 & \multirow{3}{*}{2.87E-04} & \multirow{3}{*}{4.35E+00} & \multirow{3}{*}{6.07E-03} & \multirow{3}{*}{3.88E-01} & \multirow{3}{*}{1.69E-03} \\
 & Size & 64x64 &  &  &  &  &  & 64x64 &  &  &  &  &  \\
 & Bits per cell & 32 &  &  &  &  &  & 32 &  &  &  &  &  \\
 \midrule
\textbf{Total} &  &  & \textbf{3.13E+01} & \textbf{2.79E+03} & \multicolumn{2}{r}{\textbf{4.45E+02}} & \textbf{1.24E+03} &  & \textbf{3.22E+02} & \textbf{7.21E+03} & \multicolumn{2}{r}{\textbf{1.13E+00}} & \textbf{8.12E+00} \\
\toprule \toprule
\end{tabular}
}
\begin{tablenotes}
      \item {$^*$The latency is listed as individual element. $^\dagger$S+H = Sample and Hold, IR = Input Register, OR = Output Register.}
\end{tablenotes}
\end{table*}

\subsection{Stuck Weight Offsetting}
As observed in Fig.~\ref{nonidealities_result}(d), stuck $R_{\text{ON}}/R_{\text{OFF}}$ devices lead to severe performance degradation. At 1\% stuck-at fault and above, system performance can drop below 50\% accuracy, rendering the system ineffective. In response to such degradation, we apply our proposed simplified stuck weight offsetting method. Comparing Fig.~\ref{nonidealities_result}(h) against (d), it is evident that the stuck weight offsetting method improves the average accuracy across all stuck device percentages and datasets. At 1\% stuck-at fault, the average accuracy improved by as much as 20\% for the Bonn dataset and more than 10\% for SWEC-ETHZ and CHB-MIT. The largest improvement was found for the CHB-MIT dataset at 5\% stuck-at fault, improving accuracy by 32.11\%. At higher stuck device percentages, reduced accuracy recovery is observed. This can be explained by the fact that at higher stuck device percentages, more network information cannot be recovered. Minimizing the contribution of stuck weight cannot fully retrieve the missing information, thereby leading to reduced accuracy recovery. In addition, the proposed method greatly reduces the standard deviation across patients and seeds, thanks to reduced contribution of stuck $R_{\text{ON}}/R_{\text{OFF}}$ devices to final output.

The limitation of this method lies within its inability to deal with both elements of the complementary weight being stuck $R_{\text{ON}}$ and $R_{\text{OFF}}$ simultaneously. If a positive (negative) weight is stuck $R_{\text{ON}}$ and negative (positive) weight is stuck $R_{\text{OFF}}$, stuck weight offsetting cannot provide any further adjustment to minimize the error. Meanwhile, if both weights are stuck $R_{\text{ON}}$ or $R_{\text{OFF}}$, the lost weights cannot be recovered, contributing nothing to the final output.

\begin{table*}[!t]
\centering
\caption{Performance summary and comparison of our simulated system and existing seizure detection/prediction system implementations in the literature.}\label{table:comparison_literature}
\centering
\resizebox{1\textwidth}{!}{%
\begin{tabular}{lllcccrrrrcr} \toprule \toprule
\textbf{Paper} & \textbf{Technology} & \textbf{Algorithm(s)} & \begin{tabular}[c]{@{}c@{}}\textbf{No.}\\\textbf{Channels}\end{tabular} & \begin{tabular}[c]{@{}c@{}}\textbf{Analog}\\\textbf{Front-End$^*$}\end{tabular} & \begin{tabular}[c]{@{}c@{}}\textbf{Feature}\\\textbf{Extract.$^\dagger$}\end{tabular} & \begin{tabular}[c]{@{}r@{}}\textbf{Area}\\\textbf{(mm$^2$)}\end{tabular} & \begin{tabular}[c]{@{}r@{}}\textbf{Latency}\\\textbf{(s)}\end{tabular} & \begin{tabular}[c]{@{}r@{}}\textbf{Power}\\\textbf{(mW)}\end{tabular} & \begin{tabular}[c]{@{}r@{}}\textbf{Energy}\\\textbf{(uJ)}\end{tabular} & \textbf{Pred.$^\diamond$} & \begin{tabular}[c]{@{}r@{}}\textbf{Eval.}\\\textbf{Task(s)}\end{tabular} \\ \midrule
\multicolumn{11}{c}{\textbf{ML-Based}} & \multicolumn{1}{l}{} \\ \midrule
\cite{Yoo2013} & CMOS (180nm) & \begin{tabular}[c]{@{}l@{}}BPF, \\LSVM\end{tabular} & 8 & \ding{51} & \ding{51} & 25.00 & 2.00 & N/R$^\circ$ & N/R$^\circ$ & \ding{55} & CHB-MIT \\
\cite{Altaf2013} & CMOS (180nm) & BPF, NL$-$SVM & 8 & \ding{51} & \ding{51} & 25.00 & 2.00 & N/R$^\circ$ & N/R$^\circ$ & \ding{55} & CHB-MIT \\
\cite{Lee2013} & CMOS (130nm) & NL$-$SVM & 18 & \ding{55} & \ding{51} & N/R$^\circ$ & 4.80 & N/R$^\circ$ & N/R$^\circ$ & \ding{55} & CHB-MIT \\
\cite{Chen2014} & CMOS (180nm) & \begin{tabular}[c]{@{}l@{}}FFT, \\ApEn,\\LLS\end{tabular} & 8 & \ding{51} & \ding{51} & 13.47 & 0.8 & 2.80 & \begin{tabular}[c]{@{}r@{}}2.24E\\ +03\end{tabular} & \ding{55} & In Vivo \\
\cite{BinAltaf2015} & CMOS (180nm) & \begin{tabular}[c]{@{}l@{}}BPF, \\D$^2$A$-$LSVM\end{tabular} & 16 & \ding{51} & \ding{51} & 25.0 & 1.0 & N/R$^\circ$ & N/R$^\circ$ & \ding{55} & CHB-MIT \\
\cite{BinAltaf2016} & CMOS (180nm) & \begin{tabular}[c]{@{}l@{}}BPF, \\NL$-$SVM\end{tabular} & 8 & \ding{51} & \ding{51} & 25.0 & 2.0 & 0.23 & 460.00 & \ding{55} & CHB-MIT \\
\cite{Kassiri2017} & CMOS (130nm) & FIR, PLV & 64 & \ding{51} & \ding{51} & 3.86 & N/R$^\circ$ & 1.07 & N/R$^\circ$ & \ding{51} & In Vivo \\
\cite{OLeary2018} & CMOS (130nm) & \begin{tabular}[c]{@{}l@{}}FIR, PLV/\\SE/CFC\end{tabular} & 32 & \ding{51} & \ding{51} & 7.59 & 0.25 & 0.71 & 177.50 & \ding{51} & In Vivo \\
\cite{Wang2020a} & CMOS (180nm) & \begin{tabular}[c]{@{}l@{}}DWT,\\KDE,\\SVM\end{tabular} & 8 & \ding{51} & \ding{51} & 5.83 & N/R$^\circ$ & 0.67 & N/R$^\circ$ & \ding{51} & CHB-MIT \\
\cite{Huang2020} & CMOS (40nm) & \begin{tabular}[c]{@{}l@{}}FFT,\\NL$-$SVM\end{tabular} & 14 & \ding{55} & \ding{51} & 4.50 & 0.71 & 1.90 & \begin{tabular}[c]{@{}r@{}}1.35E\\ +03\end{tabular} & \ding{55} & CHB-MIT \\
\cite{Uran2021} & CMOS (65nm) & \begin{tabular}[c]{@{}l@{}}CHT,\\XGBoost$-$DT\end{tabular} & 16 & \ding{51} & \ding{51} & 0.38 & N/R$^\circ$ & 0.40 & N/R$^\circ$ & \ding{55} & CHB-MIT, iEEG.org \\
\cite{Lin2018} & CMOS (180nm) & FFT & 1 & \ding{51} & \ding{51} & N/R$^\circ$ & N/R$^\circ$ & \ding{55}.89 & N/R$^\circ$ & \ding{55} & CHB-MIT \\
\cite{Yang2014} & CMOS (90nm) & ICA & 8 & \ding{55} & \ding{51} & 0.4 & 0.1 & \begin{tabular}[c]{@{}r@{}}8.16E \\ -02\end{tabular} & 8.16 & \ding{55} & In Vivo \\
\cite{Chen2013} & CMOS (180nm) & LLS & 1 & \ding{51} & \ding{51} & 10.41 & 0.72 & \begin{tabular}[c]{@{}r@{}}2.86E \\ -02\end{tabular} & 20.59 & \ding{55} & In Vivo \\ \midrule
\multicolumn{11}{c}{\textbf{DL-Based}} & \multicolumn{1}{l}{} \\ \midrule
\cite{Chen2018} & CMOS (65nm) & RNN & 8 & \ding{55} & \ding{55} & 10.15 & N/R$^\circ$ & 1\ding{55}.80 & N/R$^\circ$ & \ding{55} & N/R$^\circ$ \\
\cite{Daoud2018} & \begin{tabular}[c]{@{}l@{}}FPGA \\(M2GL 025-VF256)\end{tabular} & MLP & 1 & \ding{55} & \ding{51} & N/R$^\circ$ & N/R$^\circ$ & 159.70 & N/R$^\circ$ & \ding{55} & Bonn \\
\cite{Ronchini2021} & CMOS (180nm) & SNN & 1 & \ding{55} & \ding{51} & 0.15 & \begin{tabular}[c]{@{}r@{}}64.98E \\ -03\end{tabular} & \begin{tabular}[c]{@{}r@{}}5.40E \\ -03\end{tabular} & 0.35 & \ding{51} & In Vivo \\
\textbf{Ours (TDM) } & \multirow{2}{*}{\begin{tabular}[c]{@{}l@{}}\textbf{CMOS (22nm)/}\\\textbf{RRAM (BEOL) }\end{tabular}} & \multirow{2}{*}{\begin{tabular}[c]{@{}l@{}}\textbf{Manual feature }\\\textbf{extraction, }\\\textbf{CNN }\end{tabular}} & \multirow{2}{*}{\textbf{22 }} & \multirow{2}{*}{\textbf{\ding{55}}} & \multirow{2}{*}{\textbf{\ding{55}}} & \textbf{31.25} & \begin{tabular}[c]{@{}r@{}}\textbf{4.45E}\\\textbf{ -04}\end{tabular} & \begin{tabular}[c]{@{}r@{}}\textbf{2.79E }\\\textbf{ +03}\end{tabular} & \begin{tabular}[c]{@{}r@{}}\textbf{1.24E}\\\textbf{+03}\end{tabular} & \multirow{4}{*}{\textbf{\ding{51}}} & \multirow{4}{*}{\begin{tabular}[c]{@{}r@{}}\textbf{Bonn,}\\\textbf{CHB-MIT, }\\\textbf{ETHZ-}\\\textbf{SWEC }\end{tabular}}\\
\addlinespace[2ex] \textbf{Ours (Par.) } &  &  &  &  &  & \textbf{322.31} & \begin{tabular}[c]{@{}r@{}}\textbf{1.13E }\\\textbf{ -06}\end{tabular} & \begin{tabular}[c]{@{}r@{}}\textbf{7.20E }\\\textbf{ +03}\end{tabular} & \textbf{8.12} &  & \\ \bottomrule \bottomrule
\end{tabular}
}
\begin{tablenotes}
      \item {$^*$Reported power, area, and latency requirements include the analog front end/signal acquisition component. $^\dagger$Reported power, area, and latency requirements include feature extraction component(s). $^\diamond$Denotes whether systems are able to perform epileptic detection and/or prediction. $^\circ$Not reported.}
\end{tablenotes}
\end{table*}

\subsection{Power, Area, and Latency Requirements}
The following assumptions, all supported by \ac{SOTA} \ac{DL} accelerators, are made when estimating the power, area and latency requirements of our proposed memristive \ac{DL} accelerator depicted in Fig.~\ref{fig:block_diagram}, targeting a 22nm \ac{CMOS} process with device integration at the \ac{BEOL}. A memristive device has a fixed area of $100\times100$ nm$^2$~\cite{Lv2021,Yu2021} and the device read latency is $6$ ns~\cite{Zidan2018}. An ADC operating frequency is $10$ MHz~\cite{Zidan2018}, with a power consumption of $10$ mW~\cite{Zidan2018} and a device area of $1.1\times0.6$ mm$^2$~\cite{Yoshioka2005,Yu2021}. A \ac{DAC} operating frequency is $1.25$ GHz, with per unit power consumption of $6$ mW and a device area of $0.0576$ mm$^2$~\cite{Jung2008}. Other peripheral circuitry with different purposes, including the activation function~\cite{Giordano2019}, average pooling layer made up from 4-to-1 multiplexers~\cite{Li2016,Gaillardon2011}, Sample and Hold (S+H)~\cite{Shafiee2016}, subtractor~\cite{Govindarajan2020}, and adder~\cite{Ganesan2015} circuits, were listed with more detail in Table~\ref{table:power_area_latency_requirements}. 

All the peripheral components are scaled to 22nm technology by factors introduced in~\cite{Sarangi2021} and all buffers with their associated connections have energy, area and latency estimated by CACTI 7.0~\cite{CACTI7}. For all calculations, the source resistance and line resistance of $20$ $\Omega$ and $2$ $\Omega$ are used respectively. To account for RC delays within crossbars when signals are propagated, the methodology presented in~\cite{Dozortsev2018} was used, with $C_{SA}$, $T_{settling}$, and $C_{write}$ parameters from~\cite{Xu2011}. The largest total device latency was used for all devices.

In Table~\ref{table:power_area_latency_requirements}, four scenarios are considered: two where the resistance of all active (utilized) devices was fixed to $\bar{R_{\text{ON}}} \approx 10$ $k\Omega$, while considering either \ac{TDM} or parallel use of \ac{ADC}, and two where the average resistance of all active devices was assumed to be $(\bar{R_{\text{ON}}} + \bar{R_{\text{OFF}}})/2 \approx 55$ $k\Omega$, again for either \ac{TDM} or parallelized \ac{ADC}. These resistance values are representative of two weight distributions: uniform, where all weights are zero, and normal, where all weights are centered around zero. The first distribution was used to report the maximum possible power consumption of our system, and the second distribution was used to report the power consumption of a typical \ac{CNN} trained using L2-regularization. Considering the marginal impact on total power consumption, (~0.16\% and ~0.06\% for \ac{TDM} and parallelized configurations, respectively), the power of each individual trained \ac{CNN} was not determined or reported.

For all scenarios, constant operation at 0.3V per cell~\cite{Shim2020} was assumed. Neither \ac{RRAM} crossbar tiles nor peripheral circuitry was assumed to be stacked vertically. Consequently, the circuit area consumption was computed as the summation of all individual elements. Both \acp{ADC} and \acp{DAC} were assumed to operate at 6-bit resolution, as stated in Section~\ref{sec:QAT}, for the best performance with \ac{QAT}.

As can be observed in Table~\ref{table:power_area_latency_requirements}, \ac{TDM} implementations consume significantly less power than parallelized implementations due to the smaller number of required \acp{ADC}. For the worst case \ac{TDM} scenario, i.e, when all active devices are programmed to $R_{ON}^-$ with a constant 0.3V read voltage, our proposed memristive \ac{DL} accelerator has a latency of $445.22$ $\mu$s, and consumes approximately $2.79$W and $31.255$ mm$^2$ of power and area. This is fairly low power consumption for a \ac{DL} accelerator to reside on a separate chip from the neural implant, whereby the implant uses thermal energy to wirelessly communicate with the accelerator~\cite{Elansary2021}, for reduced latency.

It is noted that we have chosen to optimize the latency of our system at the cost of higher power consumption for multiple reasons. Firstly, analog crossbars which are used to perform \ac{IMC} operations, in particular \acp{VMM}, require peripheral circuitry which is power- and area-hungry. Consequently, independent of the latency of the system, when inference is being performed, a large proportion of the total system’s area and power is consumed by peripheral circuitry, registers, and buffers. While \ac{TDM} \acp{ADC} can be used to reduce the total power consumption by increasing latency, other peripheral circuits, registers, and buffers, are still required for operation. Counterintuitively, in certain instances, the energy of the system can be reduced by minimizing system latency during active operation. In other instances, the performance of the system can greatly be improved at the cost of increased power consumption.

Secondly, \ac{RRAM} devices suffer from conductance drift induced by read disturbances, which may aggregate, as the analog current is summed up along each \ac{WL} during inference~\cite{Shim2020}. To mitigate this behavior, we have constrained the absolute amplitude of \ac{BL} voltages to 0.3V and minimized the duration in which a voltage is applied to each device, i.e., latency is minimized to avoid read disturbances, and to prolong the lifespan of \ac{RRAM} devices, at the cost of increased power consumption. Lastly, as \ac{RRAM} devices are non-volatile, gating circuitry can be used to reduce the energy consumption of both \ac{TDM} and parallelized architectures, as both of our architectures have a critical delay path which is much shorter than typical signal acquisition sampling rate periods. This also allows for input buffering to be performed, so that constant operation is not required.

\subsection{Comparison to Existing Hardware Implementations}
In Table~\ref{table:comparison_literature}, we compare the performance of hardware implementations of notable epileptic seizure detection and/or prediction hardware systems in the literature. As many different evaluation tasks were used, we did not report performance metrics. Hardware implementations are broadly categorized as either \ac{ML}- or \ac{DL}-based. As can be observed, both of our implementations (reported for the $(\bar{R_{\text{ON}}} + \bar{R_{\text{OFF}}}) / 2$ scenario in Table~\ref{table:power_area_latency_requirements}) have significantly reduced inference latency, at the cost of higher power consumption, compared to traditional \ac{CMOS} and \ac{FPGA}-based implementations. It is worth noting that, most of the previous designs have not reported a complete power consumption analysis, are not capable of seizure prediction, and use fewer channels, which can lead to lower power consumption and silicon area. 

While our proposed system is not currently competitive in resource-constrained environments, it is intended to be used as a reference design for future works implementing epileptic seizure detection and prediction systems using \ac{CMOS} and memristors.
Using analog \ac{SRAM}, vertical stacking of crossbars and CMOS components, and partial sensing approaches, the power and area requirements of our simulated system could be greatly reduced. We aim to investigate these in our future research.

\section{Conclusion}\label{sec:conclusion}
We proposed a parallel \ac{CNN} architecture that can be used to perform both epileptic seizure detection and prediction rapidly. Compared to other works in literate, our architecture requires significantly fewer parameters, and demonstrates competitive performance on the University of Bonn, CHB-MIT, and SWEC-ETHZ datasets. Using emerging memristive devices and software-hardware optimization methodologies, we demonstrated, through comprehensive simulations, that our memristive \ac{DL} accelerator is capable of performing real-time operation, and consuming reasonable power in real-world conditions. We also proposed and investigated a new simplified stuck weight offsetting method to improve the robustness of our system to non-idealities. This paper sets a clear path towards the eventual circuit-level realization of a memristive epileptic seizure detection and prediction system.

\section*{Acknowledgment}
C. Lammie acknowledges the JCU DRTPS and IBM PhD Fellowship Program. M. Rahimi Azghadi acknowledges a JCU Rising Start ECR Fellowship. R. Genov and A. Amirsoleimani acknowledge the NSERC. In addition, R. Genov acknowledges the CIHR. We thank the handling editor and reviewers’ for their constructive feedback. In particular, we acknowledge the second reviewer, who provided advice on simplifying our proposed stuck weight mitigation strategy.

\bibliographystyle{IEEEtran}
\bibliography{References}

\vspace{-2em}

\begin{IEEEbiography}[{\includegraphics[width=1in,height=1.2in,clip,keepaspectratio]{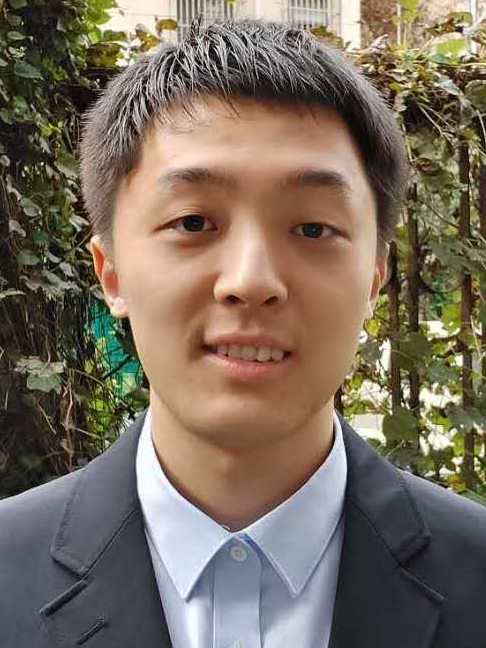}}]{Chenqi Li} 
is currently pursuing a B.A.Sc in Engineering Science, Robotics at University of Toronto, Canada. His current research interests include machine learning, computer vision, and brain-inspired computing.
\end{IEEEbiography}

\vspace{-2em}

\begin{IEEEbiography}[{\includegraphics[width=1in,height=1.2in,clip,keepaspectratio]{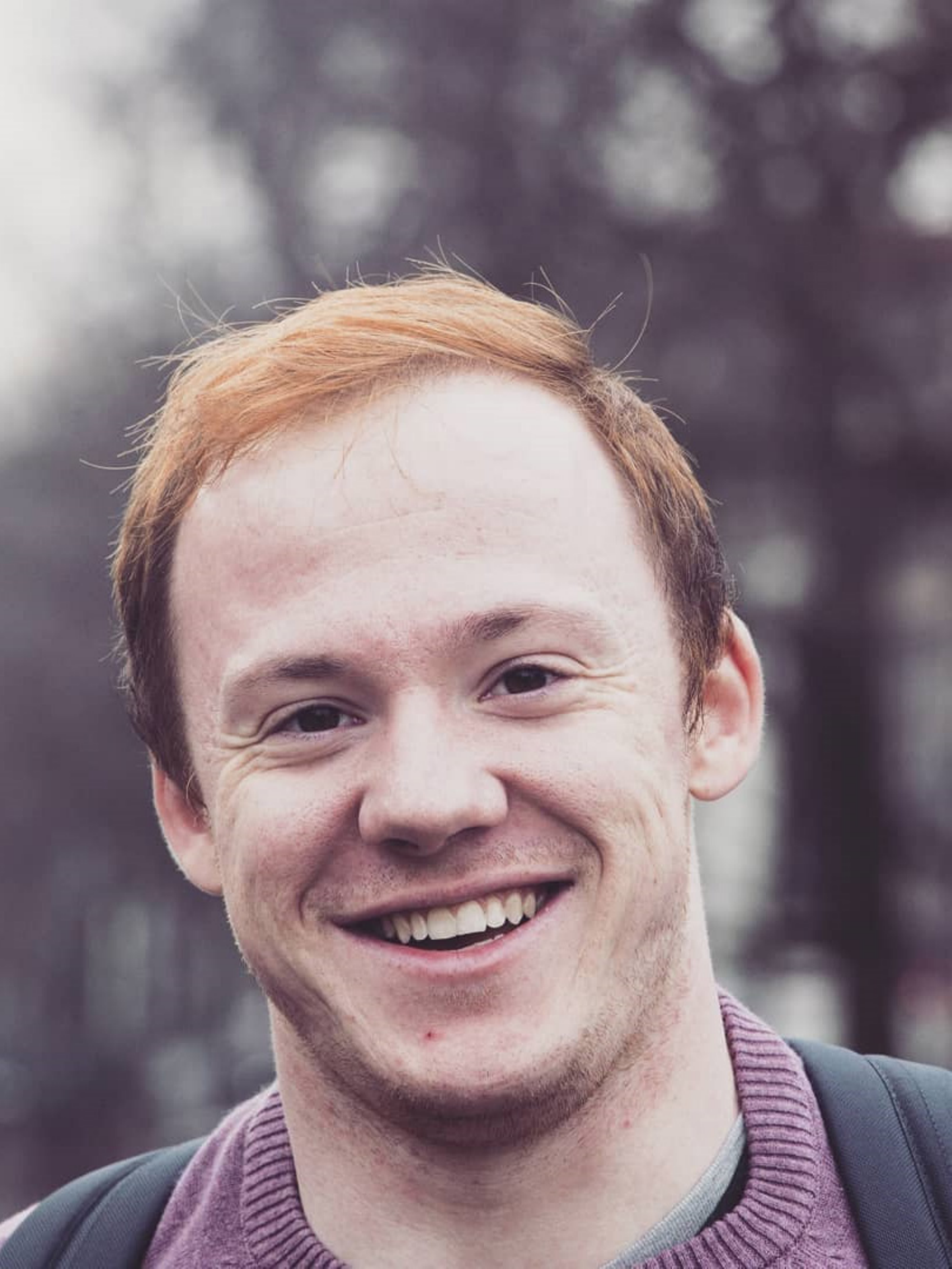}}]{Corey Lammie} 
(S'17) is currently pursuing a PhD in Computer Engineering at James Cook University (JCU), where he completed his undergraduate degrees in Electrical Engineering (Honours) and Information Technology in 2018.
His main research interests include brain-inspired computing, and the simulation and hardware implementation of Spiking Neural Networks (SNNs) and Artificial Neural Networks (ANNs) using RRAM devices and FPGAs. He has received several awards and fellowships including the intensely competitive 2020-2021 IBM international PhD Fellowship, a Domestic Prestige Research Training Program Scholarship (the highest paid PhD scholarship in Australia), the 2020 Circuits and Systems (CAS) Society Pre-Doctoral Grant, and the 2017 Engineers Australia CN Barton Medal awarded to the best undergraduate engineering thesis at JCU. Corey has served as a reviewer for several IEEE journals and conferences including IEEE Transactions on Circuits and Systems I and II, and the IEEE International Symposium on Circuits and Systems (ISCAS).
\end{IEEEbiography}

\newpage

\begin{IEEEbiography}[{\includegraphics[width=1in,height=1.2in,clip,keepaspectratio]{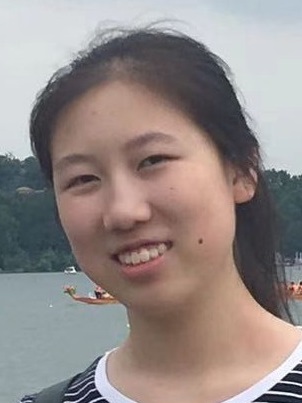}}]{Xuening Dong} 
is currently pursuing a B.A.Sc in Computer Engineering at University of Toronto, Canada. Her current research interests include machine learning, stochastic processes, and the design and simulation of memristor-based applications.
\end{IEEEbiography}

\vspace{-20em}

\begin{IEEEbiography}[{\includegraphics[width=1in,height=1.2in,clip,keepaspectratio]{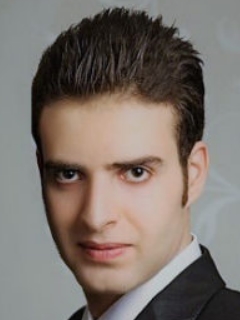}}]{Amirali Amirsoleimani} 
    (S'09--M'2017) is an assistant professor in the Department of Electrical Engineering and Computer Science at the Lassonde School of Engineering. He received his PhD in electrical and computer engineering (ECE) from University of Windsor in December 2017 and completed his postdoctoral research fellowship at the Edward S. Rogers Sr. Electrical and Computer Engineering Department at the University of Toronto in July 2021. His current research interests include application-specific processing units, in-memory computing, neuromorphic hardware design and RRAM-based accelerators for artificial intelligence. He received IEEE Larry K. Wilson award for IEEE region 7 in 2016. He was also the recipient of a best poster honourable mention award at International Joint Conference on Neural Network (IJCNN) 2017 in Alaska, USA. He is a guest editor in Frontiers in Electronics and Frontiers in Nanotechnology journals and is also serving as a reviewer for several electrical and computer engineering journals including IEEE Transactions on Circuits and Systems I (TCASI), TCAS II, TNANO, TVLSI, TED, Frontiers in Neuro-Science, Microelectronics journal, Neural Computing and Applications.
\end{IEEEbiography}

\vspace{-20em}

\begin{IEEEbiography}[{\includegraphics[width=1in,height=1.25in,clip,keepaspectratio]{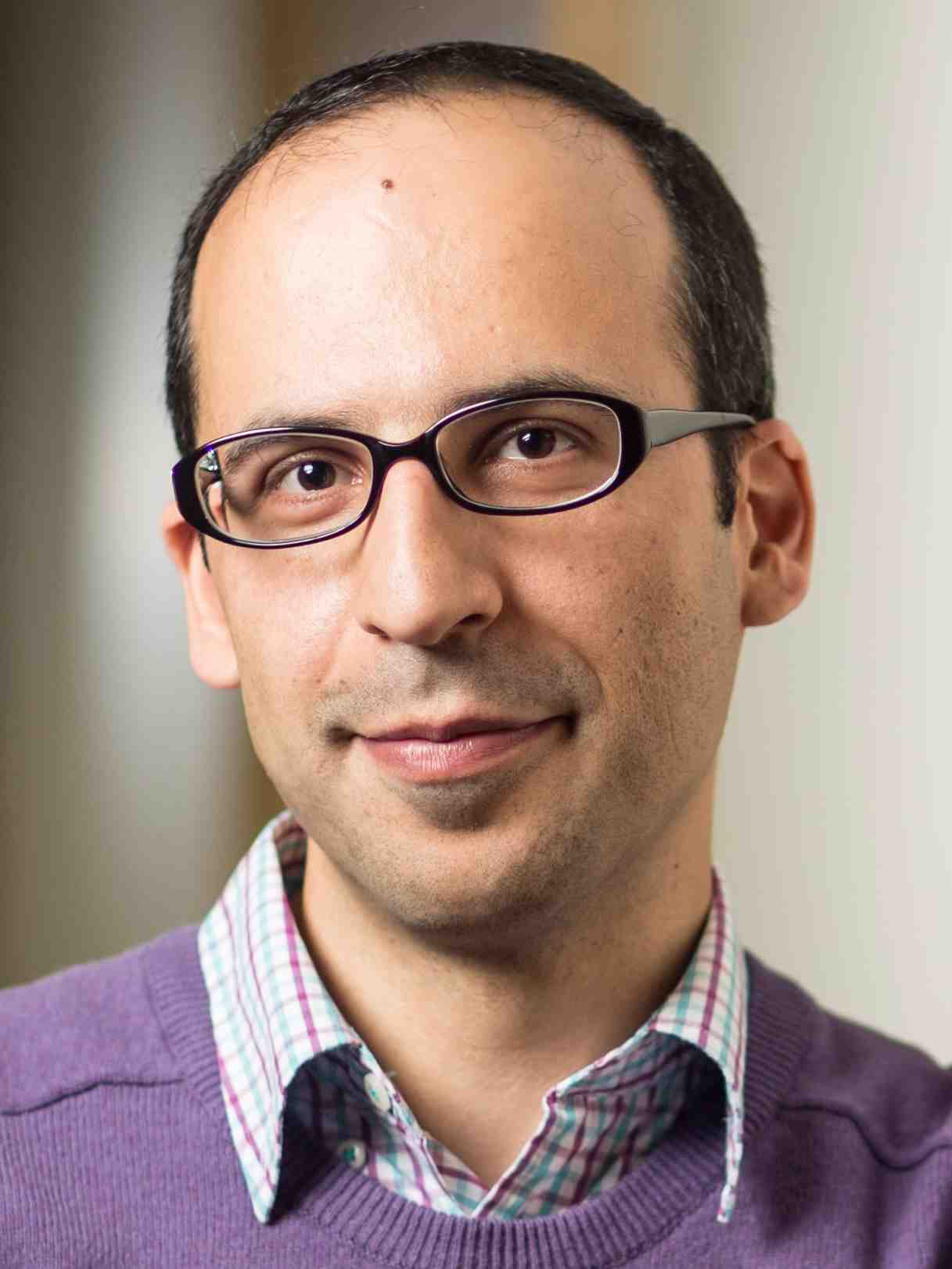}}]{Mostafa Rahimi Azghadi}
	(S'07--M'14--SM'19) completed his PhD in Electrical \& Electronic Engineering at The University of Adelaide, Australia, earning the Doctoral Research Medal, as well as the Adelaide University Alumni Medal. He is currently a senior lecturer in the College of Science and Engineering, James Cook University, Townsville, Australia, where he researches low-power and high-performance neuromorphic accelerators for neural inspired and deep learning networks for a variety of applications from agriculture to medicine. He has co-raised over \$6M in research funding from national and international resources. 
	
Dr. Rahimi was a recipient of several national and international accolades including a 2015 South Australia Science Excellence award, a 2016 Endeavour Research Fellowship, a 2017 Queensland Young Tall Poppy Science Award, a 2018 JCU Rising Star ECR Leader Fellowship, a 2019 Fresh Science Queensland Finalist, and a 2020 JCU Award for Excellence in Innovation and Change. Dr Rahimi is a senior member of the IEEE and a TC member of Neural Systems and Applications of the circuit and system society. He serves as an associate editor of Frontiers in Neuromorphic Engineering and IEEE Access.
\end{IEEEbiography}

\newpage

\begin{IEEEbiography}[{\includegraphics[width=1in,height=1.2in,clip,keepaspectratio]{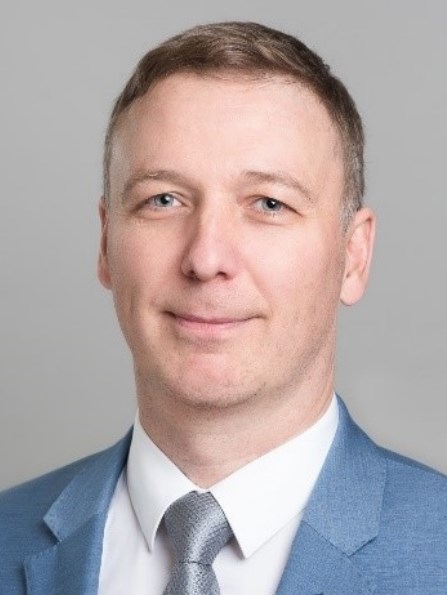}}]{Roman Genov} 
    (S’96–M’02–SM’11) received the B.S. degree in Electrical Engineering from Rochester Institute of Technology, NY in 1996 and the M.S.E. and Ph.D. degrees in Electrical and Computer Engineering from Johns Hopkins University, Baltimore, MD in 1998 and 2003 respectively.
    
He is currently a Professor in the Department of Electrical and Computer Engineering at the University of Toronto, Canada, where he is a member of Electronics Group and Biomedical Engineering Group and the Director of Intelligent Sensory Microsystems Laboratory. Dr. Genov’s research interests are primarily in analog integrated circuits and systems for energy-constrained biological, medical, and consumer sensory applications.

Dr. Genov is a co-recipient of Jack Kilby Award for Outstanding Student Paper at IEEE International Solid-State Circuits Conference, Best Paper Award of IEEE Transactions on Biomedical Circuits and Systems, Best Paper Award of IEEE Biomedical Circuits and Systems Conference, Best Student Paper Award of IEEE International Symposium on Circuits and Systems, Best Paper Award of IEEE Circuits and Systems Society Sensory Systems Technical Committee, Brian L. Barge Award for Excellence in Microsystems Integration, MEMSCAP Microsystems Design Award, DALSA Corporation Award for Excellence in Microsystems Innovation, and Canadian Institutes of Health Research Next Generation Award. He was a Technical Program Co-chair at IEEE Biomedical Circuits and Systems Conference, a member of IEEE European Solid-State Circuits Conference Technical Program Committee, and a member of IEEE International Solid-State Circuits Conference International Program Committee. He was also an Associate Editor of IEEE TCAS II and IEEE Signal Processing Letters, as well as a Guest Editor for IEEE JSSC. Currently he is an Associate Editor of IEEE Transactions on Biomedical Circuits and Systems.
\end{IEEEbiography}

\end{document}